\documentstyle[11pt]{article}
\newcommand{\rf}[1]{(\ref{#1})}
\newcommand{\bea}{\begin{eqnarray}}
\newcommand{\eea}{\end{eqnarray}}

\newcommand{\G}{\Gamma}

\renewcommand{\L}{\Lambda}

\renewcommand{\a}{\alpha}
\newcommand{\n}{\nu}
\newcommand{\m}{\mu}

\newcommand{\k}{\kappa}
\newcommand{\del}{\delta}

\newcommand{\ra}{\right\rangle}
\newcommand{\la}{\left\langle}

\newcommand{\cD}{{\cal D}}

\newcommand{\cs}{{\rm cs}}

\def\void{}
\def\labelmark{}

\newenvironment{formula}[1]{\def\labelname{#1}
\ifx\void\labelname\def\junk{\begin{displaymath}}
\else\def\junk{\begin{equation}\label{\labelname}}\fi\junk}%
{\ifx\void\labelname\def\junk{\end{displaymath}}
\else\def\junk{\end{equation}}\fi\junk\labelmark\def\labelname{}}

{\ifx\void\labelname\def\junk{\end{array}\end{displaymath}}
\else\def\junk{\end{array}\right.\end{equation}}
\fi\junk\labelmark\def\labelname{}\def\junk{}
}

\newcommand{\beq}{\begin{formula}}
\newcommand{\eeq}{\end{formula}}
\newcommand{\beqv}{\begin{formula}{}}

\begin{document}
\topmargin 0pt
\oddsidemargin 5mm
\headheight 0pt
\headsep 0pt
\topskip 9mm

\hfill    NBI-HE-95-23

\hfill June 1995

\begin{center}
\vspace{24pt}
{\large \bf The classical sphaleron transition rate exists and is equal to
\boldmath $1.1(\alpha_{\bf w} T)^4$}

\vspace{24pt}

{\sl J. Ambj\o rn} and {\sl A. Krasnitz}

 The Niels Bohr Institute\\
Blegdamsvej 17, DK-2100 Copenhagen \O , Denmark\\

\end{center}
\vspace{24pt}

%\addtolength{\baselineskip}{0.20\baselineskip}
\vfill

\begin{center}
{\bf Abstract}
\end{center}

\vspace{12pt}

\noindent
Results of a large scale numerical simulation show that
the high temperature Chern-Simons number diffusion rate
in the electroweak theory has a classical limit $\G = \k (\a_w T)^4$, where
$\k = 1.09\pm 0.04$ and $\a_w$ is the weak fine structure constant.

\vfill

\newpage

\section{Introduction}

Topology-changing transitions play an important role in the electroweak
theory since they are accompanied by baryon-number violating processes
\cite{thooft}.
At zero temperature these processes can only appear as
true quantum effects since the energy of the sphaleron configuration
is a barrier separating two gauge equivalent vacua \cite{sphaleron}.
At temperatures far above the sphaleron energy one would expect
that this barrier height,
as well as the concept of vacuum to vacuum amplitude,
are irrelevant for the transition rate.
Unfortunately there presently is no way to calculate
analytically the transition rate at these high temperatures.
{}From a numerical point of view the situation is not any better if we
insist on a fully quantum treatment, since the only
non-perturbative tool available is lattice gauge theories
in the imaginary time formalism. Real time transition rates
are not easily extracted in this formalism.

The object of our study is the topological charge
\beq{1.1}
B(t) = \frac{1}{32 \pi^2}\int^t_0 \int d^3x\,
F^a_{\m\n}\tilde{F}^{a \m\n}= N_{\cs}(t)-N_{\cs}(0),
\eeq
where the Chern-Simons number
\beq{1.1b}
N_{\cs}(t) \equiv \frac{1}{32\pi^2} \,\epsilon_{ijk} \int d^3x
\left(F^a_{ij} A^a_k - \frac{1}{3} \epsilon^{abc} A^a_iA^b_jA^c_k\right).
\eeq
The theory has a periodic (with period 1) potential in the
$N_{\rm cs}$ direction. This periodicity is due to the symmetry under large
gauge transformations, changing $N_{\rm cs}$ by an integer. There also are
dynamical processes whose result for a field configuration is the same as a
large gauge transformation. These processes have an integer topological
charge. Due to nonlinear interactions of $N_{\rm cs}$ with other degrees
of freedom of the theory it is expected that for large times $t$ $B(t)$ is
a random walk in a periodic $N_{\rm cs}$ potential
\beq{1.1x}
\la B^2(t)\ra_T=\G Vt,
\eeq
where $V$ is the space volume,
by $\la\ra_T$ we mean average over the canonical ensemble, and $\G$, the
diffusion constant per unit volume, is the transition rate we are interested in
\cite{khls}.

At high temperatures  processes with an integer topological charge occur
predominantly by classically allowed thermal activation rather than by quantum
tunneling. Some time
ago it was suggested \cite{grs} to exploit this property and to determine
$\G$ entirely
within the classical theory. This is done by solving {\it classical} equations
of motion for the fields and averaging the resulting $B^2(t)$ over the
{\it classical} canonical ensemble. Behind this approach is the expectation
that the topology-changing transitions mostly involve field fluctuations with
long wavelengths and large magnitudes (hence many quanta). Such fluctuations
are well described by classical statistical mechanics. An argument usually
given to support this picture is as follows. Consider a classically allowed
process with $B(t)=1$, whereby the field configuration evolves from the
vicinity of one minimum in the $N_{\rm cs}$ potential to the vicinity of
another
one. If a typical linear space-time size of the process is $r$, the system
will cross an energy barrier of height $E\sim 1/\alpha r$, where
$\alpha=g^2/4\pi$. While large $r$ are favored energetically,
at high temperatures $r$ is unlikely to exceed the inverse
of the magnetic screening mass $m_{\rm mag}\sim\alpha T$. We then expect that
processes with $r\sim 1/\alpha T$ give the most important contribution to
$\Gamma$. In the electroweak
theory, where $\alpha_{\rm w}\approx 1/30$, these $r$ are much larger than the
thermal
wavelength, {\it i.e.,} fluctuations of size $r$ can be treated classically.
In order to estimate the transition rate, the Boltzmann factor associated with
the barrier, $\exp(-E/T)\sim 1$, is divided by the space-time volume $r^4$
of the process, to give \cite{am}
\beq{1.1y}
\Gamma=\k (\alpha T)^4,\label{esti}\eeq
where $\k$ is a dimensionless constant.
This estimate should be valid at temperatures far above the electroweak
transition
scale. We also expect that at such high temperatures the mass scales of the
Higgs sector are irrelevant, and that $\Gamma$ can be reliably obtained from
a pure Yang-Mills theory.

The classical approximation offers an enormous simplification in computing
real-time correlation functions. However,
it is well known that classical field theory at finite temperature suffers
from the Rayleigh-Jeans ultraviolet divergence. This means that, upon cutoff
regularization, classical thermodynamic average of a generic observable
will be sensitive to thermal fluctuations at the cutoff scale $\Lambda$
\cite{bms}.
The classical rate $\Gamma$, if it is to be trusted, must not show such
sensitivity. This is indeed what we expect: considerations leading to
the estimate (\ref{esti}) may be repeated for a purely classical theory which
also possesses a magnetic mass $m_{\rm mag}\sim\alpha T$. In the quantum
case, we required separation between the magnetic and the thermal scales:
$\alpha T\ll T$. Likewise, in the classical theory we should require separation
between the magnetic and the cutoff scales: $\alpha T\ll \Lambda$. These
two conditions make it plausible that $\G$ is not strongly affected by either
quantum corrections or the classical cutoff.

On the other hand,
it is not clear how the $\Gamma$ independence of $\Lambda$ is technically
possible: after
all, $\dot B(t)$, the topological charge per unit time, is an ultraviolet
divergent quantity whose classical standard deviation is given perturbatively
by
\beq{1.3b}
\la (\dot B )^2 \ra_T =
\left(\frac{1}{8\pi^2}\right)^2
\la  \left(\int d^3x E^a_i(x)B^a_i(A(x)) \right)^2 \ra_T \sim
T^2 V \L^3,
\eeq
where $E^a_i$ and $B^a_i$ are, respectively, color electric and magnetic
fields,
and $V$ is the space volume.
One then naturally expects the Rayleigh-Jeans divergency to show up in
$\la B^2(t)\ra_T$ itself.

The key point, confirmed by the numerical simulations in the following, is that
the UV divergency shows up in $\la B^2(t)\ra_T$ in a way that does not affect
$\G$. To see how this could be anticipated, consider the structure of
$\la B^2(t)\ra_T$ in some more detail. The classical motion of $N_{\rm cs}$ can
be thought of as consisting of two pieces: ``true'' thermal fluctuations
and a random walk between gauge equivalent vacua.
If the
theory did not have an infinite series of degenerate vacua, $N_{\cs}$ would
simply fluctuate around zero, as it does in an Abelian theory.
The properties of these thermal fluctuations can be studied perturbatively.
As discussed in the Appendix,
in an Abelian theory, where no random walk of $N_{\rm cs}$ is possible,
$\la B^2(t)\ra_T$ is bounded
\beq{1.3d}
\la B^2(t)\ra_T^{{\rm Abelian}}\leq
2\la (N_{\cs})^2 \ra_T^{{\rm Abelian}}
\eeq
and at large $t$ fluctuates around the average value
\beq{1.3dd}
\nu\equiv\la (N_{\cs})^2 \ra_T^{{\rm Abelian}}\approx 0.00228
{{\left(\a T\right)^2}\over a} V,
\eeq
diverging, as expected, linearly with the lattice cutoff $\L=1/a$.
In the nonabelian
theory the random walk of $N_{\rm cs}$ is superimposed with this thermal
background. A reasonable approximation to the nonabelian $\la B^2(t)\ra_T$
is then
\beq{1.3e}
\la B^2(t) \ra_T = c\nu  + \G V t,
\eeq
where $c\approx 3$, assuming that thermal
fluctuations of $N_{\rm cs}$ are exactly those for three independent copies
of an Abelian theory.

If we let $\L \to \infty$ for fixed $t$ the r.h.s. of (\ref{1.3e}) diverges.
On the other hand, for a fixed cutoff \rf{1.3e} is well approximated by
(\ref{1.1x}) for large enough $t$ and allows unambiguously to extract $\G$.
We will show that
numerical simulations, using a classical thermal ensemble
with a lattice cutoff $\L = 1/a$, point
to a $\G$ independent of $a$ for sufficiently small $a$.
On dimensional grounds this means that \rf{1.1y} holds classically in the
continuum limit, with $\k$ of \rf{1.1y} being
a {\it  non-perturbative constant of the classical
theory}. For the gauge group $SU(2)$ we find $\k = 1.09 \pm 0.04$.
This is the central result of our work.

The ultraviolet insensitivity of $\G$ extracted from \rf{1.3e}
might be linked to the topological
aspect of the special field tensor combination
$F_{\m\n}^a \tilde{F}^{a \m\n}$ in non-abelian field theories.
Replacing $F_{\m\n}^a \tilde{F}^{a \m\n}$ by a generic bilinear of field
tensors with zero average would yield a UV divergent transition rate.
We illustrate this point by measuring in addition the
quantity $\eta(t)$ whose diffusion rate is related to the shear viscosity of
the gluon plasma \cite{wyld}:
\beq{1.5}
\eta(t)  = \int_0^t dt \int d^3 x \left(T_{11}-T_{22}\right),
\eeq
where $T_{ij}$ are components of the energy-momentum tensor.

\section{The model and the numerical method}

The philosophy behind the classical measurements of real-time quantities
has been presented in detail
in a number of articles  \cite{jan,alex1}.
Here we only concentrate on the essential technical points.
We work in the temporal gauge. The classical dynamics of $SU(2)$ Yang-Mills
theory on the lattice is given by Kogut-Susskind Hamiltonian
\beq{2.1}
H={1\over 2}\sum_l E_l^\alpha E_l^\alpha
+ \sum_\Box\left(1-{1\over 2}{\rm Tr}U_\Box\right).
\eeq
The variable set consists of SU(2) matrices $U_l$ and lattice analogs of
electric field $E_l$, both residing on the links of a cubic lattice.
We shall also use notation $l=j,n$ for a link emanating from site $j$
in positive direction $n$.
The second term in (\ref{2.1}) is the standard plaquette term, representing the
color magnetic energy. Every variable $v$ evolves in time according to its
canonical equation $\dot v=\{H,v\}$, derived using Poisson brackets between
the independent variables:
\beq{2.2}
\{E^\alpha_l,E^\beta_{l'}\}=2\delta_{ll'}\epsilon^{\alpha\beta\gamma}E^\gamma_l;
\ \ \{E^\alpha_l,U_{l'}\}=-i\delta_{ll'}U_l\sigma^\alpha;
\ \ \{U_l,U_{l'}\}=0
\eeq
(no summation over repeated $l$; $\sigma^\alpha$ are the Pauli matrices).
The time evolution is subject to the set of three Gauss' constraints per site
\beq{2.3}
\sum_n\left(\overline{E}^\alpha_{j,n}-E^\alpha_{j-n,n}\right)=0,
\eeq
where
$$\overline{E}^\alpha_{j,n}\equiv {1\over 2}E^\beta_{j,n}{\rm Tr}
\left(\sigma^\alpha U_{j,n}\sigma^\beta U_{j,n}^\dagger\right).$$
These constraints
are to be imposed on initial conditions. We use the leapfrog algorithm to
integrate the equations of motion, performing {\it exact} integration of the
$E$
equations for fixed $U$ and of the $U$ equations for fixed $E$. In this way
the Gauss' laws are obeyed exactly, independently of the time step, and the
only
source of spurious static charge is the limited computer accuracy.

The increment of Chern-Simons number is computed as in (\ref{1.1}) with
the topological charge per unit time approximated by
\beq{2.4}
\dot N_{\rm cs}={i\over{32\pi^2}}\sum_{j,n}\left(\overline{E}^\alpha_{j,n}
+E^\alpha_{j-n,n}\right)
\sum_{\Box_{j,n}}{\rm Tr}\left(U_{\Box_{j,n}}\sigma^\alpha\right),
\eeq
where $\Box_{j,n}$ denotes plaquettes in the plane perpendicular to
direction $n$,
running counterclockwise in that plane, and beginning and ending at site $j$.

Our purpose is to study the average of $B^2(t)$ ({\it cf} (\ref{1.1})) over the
thermal ensemble of initial field configurations. Due to the presence of
first-class constraints (Gauss' laws) this point is nontrivial. The initial
configurations must satisfy the constraints and be thermally distributed in the
reduced phase space of physical, gauge-invariant degrees of freedom. This goal
is achieved by using the constraint-respecting Langevin algorithm. The set
of Langevin equations reads ($\beta$ is the inverse temperature in lattice
units)
\beq{2.5}
\dot U_{j,k}=\{H, U_{j,k}\}+\left(\Gamma_{j,mn}(t)-\beta\{T_{j,mn},H\}\right)
\{T_{j,mn},U_{j,k}\}; \ \ \dot E^\alpha_l=\{H,E^\alpha_l\},
\eeq
where, for fixed $j$, $T_{j,mn}\equiv \sqrt{\gamma}\overline{E}^\alpha_{j,m}
\overline{E}^\alpha_{j,n}$ (assuming for definitness $m\leq n$),
and $\Gamma_{j,mn}(t)$ is a random variable with zero average and correlation
$$\langle\Gamma_{j,mn}(t)\Gamma_{j',m'n'}(t')\rangle=2\delta(t-t')\delta_{jj'}
\delta_{mn,m'n'}.$$
We are free to choose the friction coefficient $\gamma>0$. This freedom can be
used to optimize the algorithm performance. Evolution generated by (\ref{2.5})
is not to be confused with real-time evolution according to canonical equations
of motion: the sole purpose of (\ref{2.5}) is to bring the system in question
to thermal equilibrium at temperature $1/\beta$.
The set of equations  (\ref{2.5}) is integrated numerically using a
modified leapfrog
algorithm, which obeys the Gauss' laws as accurately as a computer arithmetic
allows.
For a more detailed description of the algorithm we refer to \cite{alex}.

Results of simulation must eventually be converted from lattice units to
the physical ones in the continuum. To do so it is convenient to parametrize
the link matrices as $U_l=\exp(i a\sigma^\alpha A^\alpha_l/2)$ where $a$ is the
lattice spacing and $A^\alpha_l$ is the gauge potential. It follows from the
equation of motion for $U_l$ that $E^\alpha_l=i{\rm Tr}(\sigma^\alpha
U^\dagger_l\dot U_l)/2$. Therefore for small $a$ (\ref{2.1}) takes form
\beq{2.6}
H\approx {a^2\over 8}\sum_{j,n}\left(\dot A^\alpha_{j,n}\dot A^\alpha_{j,n}
+a^2 B^\alpha_{j,n}B^\alpha_{j,n}\right),
\eeq
where $B^\alpha_{j,n}$ is a chromo-magnetic field. Comparing (\ref{2.6}) to the
standard continuum Hamiltonian
$$ H_C={1\over{2 g^2}}\int d^3x \left(\dot A^\alpha_i\dot A^\alpha_i +
B^\alpha_i B^\alpha_i\right)$$
we find the correspondence between the time and the temperature expressed in
lattice ($L$) and continuum ($C$) units:
\beq{2.7}
t_C=at_L; \ \ T_C={4\over{g^2 a}} T_L. \eeq
Taking into account that $N_{\rm cs}$ itself is a dimensionless quantity,
\rf{1.1y} on the lattice takes form $\G_L=\k (T_L/\pi)^4$, where the
lattice rate $\G_L$ is per unit cell.

\section{Results}

We performed series of simulations using methods described in the previous
section for three values of the inverse lattice temperature $\beta=10,12,14$
on lattices of equal size $L_L$ and periodic boundary conditions in all
spatial directions. The value of $L_L$ ranged between 8 and 32. Between 20
and 60 statistically independent initial configurations were produced for
every $\beta,L_L$ pair (60 for the largest volumes, used for the estimate of
$\kappa$). At low temperatures considered the system is only weakly nonlinear,
and we expect an average energy per physical degree of freedom to be close to
$1/2\beta$. This is indeed what we observe: the average chromo-electric
energy per site was found to be $3/\beta$ with a three-digit accuracy.
Our measurement results for the topological charge density are also close to
the perturbative estimate \rf{5.3}.

Every independent initial configuration was let evolve in real time for
5000 time units. We used the integration time step of 0.05 for the classical
equations of motion. From our experience in earlier work \cite{jan} we know
that
reducing the time step beyond this value has negligible impact on the
diffusion of Chern-Simons number.
During the real-time evolution the value of $B(t)$, determined by
numerically integrating the lattice topological charge per unit time,
was recorded with intervals between 0.5 and 1 time unit. In the same way,
we recorded the time history of $\eta(t)$. Thus our results for
$\langle B^2(t)\rangle_T$, discussed in the following, represent
averaging both over the canonical ensemble of initial configurations and within
every individual real-time trajectory. The same averaging method was used
in earlier work on a low-dimensional model \cite{alex1}.

\begin{figure}
% GNUPLOT: LaTeX picture
\setlength{\unitlength}{0.240900pt}
\ifx\plotpoint\undefined\newsavebox{\plotpoint}\fi
\sbox{\plotpoint}{\rule[-0.200pt]{0.400pt}{0.400pt}}%
\begin{picture}(1800,1080)(0,0)
\font\gnuplot=cmr10 at 10pt
\gnuplot
\sbox{\plotpoint}{\rule[-0.200pt]{0.400pt}{0.400pt}}%
\put(220.0,113.0){\rule[-0.200pt]{365.204pt}{0.400pt}}
\put(220.0,113.0){\rule[-0.200pt]{0.400pt}{227.410pt}}
\put(220.0,113.0){\rule[-0.200pt]{4.818pt}{0.400pt}}
\put(198,113){\makebox(0,0)[r]{0}}
\put(1716.0,113.0){\rule[-0.200pt]{4.818pt}{0.400pt}}
\put(220.0,248.0){\rule[-0.200pt]{4.818pt}{0.400pt}}
\put(198,248){\makebox(0,0)[r]{0.1}}
\put(1716.0,248.0){\rule[-0.200pt]{4.818pt}{0.400pt}}
\put(220.0,383.0){\rule[-0.200pt]{4.818pt}{0.400pt}}
\put(198,383){\makebox(0,0)[r]{0.2}}
\put(1716.0,383.0){\rule[-0.200pt]{4.818pt}{0.400pt}}
\put(220.0,518.0){\rule[-0.200pt]{4.818pt}{0.400pt}}
\put(198,518){\makebox(0,0)[r]{0.3}}
\put(1716.0,518.0){\rule[-0.200pt]{4.818pt}{0.400pt}}
\put(220.0,652.0){\rule[-0.200pt]{4.818pt}{0.400pt}}
\put(198,652){\makebox(0,0)[r]{0.4}}
\put(1716.0,652.0){\rule[-0.200pt]{4.818pt}{0.400pt}}
\put(220.0,787.0){\rule[-0.200pt]{4.818pt}{0.400pt}}
\put(198,787){\makebox(0,0)[r]{0.5}}
\put(1716.0,787.0){\rule[-0.200pt]{4.818pt}{0.400pt}}
\put(220.0,922.0){\rule[-0.200pt]{4.818pt}{0.400pt}}
\put(198,922){\makebox(0,0)[r]{0.6}}
\put(1716.0,922.0){\rule[-0.200pt]{4.818pt}{0.400pt}}
\put(220.0,1057.0){\rule[-0.200pt]{4.818pt}{0.400pt}}
\put(198,1057){\makebox(0,0)[r]{0.7}}
\put(1716.0,1057.0){\rule[-0.200pt]{4.818pt}{0.400pt}}
\put(220.0,113.0){\rule[-0.200pt]{0.400pt}{4.818pt}}
\put(220,68){\makebox(0,0){0}}
\put(220.0,1037.0){\rule[-0.200pt]{0.400pt}{4.818pt}}
\put(523.0,113.0){\rule[-0.200pt]{0.400pt}{4.818pt}}
\put(523,68){\makebox(0,0){5}}
\put(523.0,1037.0){\rule[-0.200pt]{0.400pt}{4.818pt}}
\put(826.0,113.0){\rule[-0.200pt]{0.400pt}{4.818pt}}
\put(826,68){\makebox(0,0){10}}
\put(826.0,1037.0){\rule[-0.200pt]{0.400pt}{4.818pt}}
\put(1130.0,113.0){\rule[-0.200pt]{0.400pt}{4.818pt}}
\put(1130,68){\makebox(0,0){15}}
\put(1130.0,1037.0){\rule[-0.200pt]{0.400pt}{4.818pt}}
\put(1433.0,113.0){\rule[-0.200pt]{0.400pt}{4.818pt}}
\put(1433,68){\makebox(0,0){20}}
\put(1433.0,1037.0){\rule[-0.200pt]{0.400pt}{4.818pt}}
\put(1736.0,113.0){\rule[-0.200pt]{0.400pt}{4.818pt}}
\put(1736,68){\makebox(0,0){25}}
\put(1736.0,1037.0){\rule[-0.200pt]{0.400pt}{4.818pt}}
\put(220.0,113.0){\rule[-0.200pt]{365.204pt}{0.400pt}}
\put(1736.0,113.0){\rule[-0.200pt]{0.400pt}{227.410pt}}
\put(220.0,1057.0){\rule[-0.200pt]{365.204pt}{0.400pt}}
\put(45,585){\makebox(0,0){$\langle B^2(t)\rangle$}}
\put(978,23){\makebox(0,0){{\large t}}}
\put(220.0,113.0){\rule[-0.200pt]{0.400pt}{227.410pt}}
\put(220,113){\raisebox{-.8pt}{\makebox(0,0){$\Diamond$}}}
\put(281,346){\raisebox{-.8pt}{\makebox(0,0){$\Diamond$}}}
\put(341,372){\raisebox{-.8pt}{\makebox(0,0){$\Diamond$}}}
\put(402,415){\raisebox{-.8pt}{\makebox(0,0){$\Diamond$}}}
\put(463,450){\raisebox{-.8pt}{\makebox(0,0){$\Diamond$}}}
\put(523,485){\raisebox{-.8pt}{\makebox(0,0){$\Diamond$}}}
\put(584,518){\raisebox{-.8pt}{\makebox(0,0){$\Diamond$}}}
\put(644,549){\raisebox{-.8pt}{\makebox(0,0){$\Diamond$}}}
\put(705,578){\raisebox{-.8pt}{\makebox(0,0){$\Diamond$}}}
\put(766,606){\raisebox{-.8pt}{\makebox(0,0){$\Diamond$}}}
\put(826,632){\raisebox{-.8pt}{\makebox(0,0){$\Diamond$}}}
\put(887,659){\raisebox{-.8pt}{\makebox(0,0){$\Diamond$}}}
\put(948,685){\raisebox{-.8pt}{\makebox(0,0){$\Diamond$}}}
\put(1008,711){\raisebox{-.8pt}{\makebox(0,0){$\Diamond$}}}
\put(1069,735){\raisebox{-.8pt}{\makebox(0,0){$\Diamond$}}}
\put(1130,760){\raisebox{-.8pt}{\makebox(0,0){$\Diamond$}}}
\put(1190,784){\raisebox{-.8pt}{\makebox(0,0){$\Diamond$}}}
\put(1251,809){\raisebox{-.8pt}{\makebox(0,0){$\Diamond$}}}
\put(1312,833){\raisebox{-.8pt}{\makebox(0,0){$\Diamond$}}}
\put(1372,857){\raisebox{-.8pt}{\makebox(0,0){$\Diamond$}}}
\put(1433,881){\raisebox{-.8pt}{\makebox(0,0){$\Diamond$}}}
\put(1493,904){\raisebox{-.8pt}{\makebox(0,0){$\Diamond$}}}
\put(1554,929){\raisebox{-.8pt}{\makebox(0,0){$\Diamond$}}}
\put(1615,951){\raisebox{-.8pt}{\makebox(0,0){$\Diamond$}}}
\put(1675,975){\raisebox{-.8pt}{\makebox(0,0){$\Diamond$}}}
\put(1736,999){\raisebox{-.8pt}{\makebox(0,0){$\Diamond$}}}
\put(220,113){\usebox{\plotpoint}}
\put(210.0,113.0){\rule[-0.200pt]{4.818pt}{0.400pt}}
\put(210.0,113.0){\rule[-0.200pt]{4.818pt}{0.400pt}}
\put(281.0,345.0){\rule[-0.200pt]{0.400pt}{0.482pt}}
\put(271.0,345.0){\rule[-0.200pt]{4.818pt}{0.400pt}}
\put(271.0,347.0){\rule[-0.200pt]{4.818pt}{0.400pt}}
\put(341.0,372.0){\usebox{\plotpoint}}
\put(331.0,372.0){\rule[-0.200pt]{4.818pt}{0.400pt}}
\put(331.0,373.0){\rule[-0.200pt]{4.818pt}{0.400pt}}
\put(402.0,414.0){\rule[-0.200pt]{0.400pt}{0.482pt}}
\put(392.0,414.0){\rule[-0.200pt]{4.818pt}{0.400pt}}
\put(392.0,416.0){\rule[-0.200pt]{4.818pt}{0.400pt}}
\put(463.0,449.0){\rule[-0.200pt]{0.400pt}{0.482pt}}
\put(453.0,449.0){\rule[-0.200pt]{4.818pt}{0.400pt}}
\put(453.0,451.0){\rule[-0.200pt]{4.818pt}{0.400pt}}
\put(523.0,484.0){\rule[-0.200pt]{0.400pt}{0.723pt}}
\put(513.0,484.0){\rule[-0.200pt]{4.818pt}{0.400pt}}
\put(513.0,487.0){\rule[-0.200pt]{4.818pt}{0.400pt}}
\put(584.0,516.0){\rule[-0.200pt]{0.400pt}{0.723pt}}
\put(574.0,516.0){\rule[-0.200pt]{4.818pt}{0.400pt}}
\put(574.0,519.0){\rule[-0.200pt]{4.818pt}{0.400pt}}
\put(644.0,547.0){\rule[-0.200pt]{0.400pt}{0.964pt}}
\put(634.0,547.0){\rule[-0.200pt]{4.818pt}{0.400pt}}
\put(634.0,551.0){\rule[-0.200pt]{4.818pt}{0.400pt}}
\put(705.0,576.0){\rule[-0.200pt]{0.400pt}{0.964pt}}
\put(695.0,576.0){\rule[-0.200pt]{4.818pt}{0.400pt}}
\put(695.0,580.0){\rule[-0.200pt]{4.818pt}{0.400pt}}
\put(766.0,604.0){\rule[-0.200pt]{0.400pt}{0.964pt}}
\put(756.0,604.0){\rule[-0.200pt]{4.818pt}{0.400pt}}
\put(756.0,608.0){\rule[-0.200pt]{4.818pt}{0.400pt}}
\put(826.0,630.0){\rule[-0.200pt]{0.400pt}{1.204pt}}
\put(816.0,630.0){\rule[-0.200pt]{4.818pt}{0.400pt}}
\put(816.0,635.0){\rule[-0.200pt]{4.818pt}{0.400pt}}
\put(887.0,656.0){\rule[-0.200pt]{0.400pt}{1.445pt}}
\put(877.0,656.0){\rule[-0.200pt]{4.818pt}{0.400pt}}
\put(877.0,662.0){\rule[-0.200pt]{4.818pt}{0.400pt}}
\put(948.0,682.0){\rule[-0.200pt]{0.400pt}{1.445pt}}
\put(938.0,682.0){\rule[-0.200pt]{4.818pt}{0.400pt}}
\put(938.0,688.0){\rule[-0.200pt]{4.818pt}{0.400pt}}
\put(1008.0,708.0){\rule[-0.200pt]{0.400pt}{1.445pt}}
\put(998.0,708.0){\rule[-0.200pt]{4.818pt}{0.400pt}}
\put(998.0,714.0){\rule[-0.200pt]{4.818pt}{0.400pt}}
\put(1069.0,731.0){\rule[-0.200pt]{0.400pt}{1.927pt}}
\put(1059.0,731.0){\rule[-0.200pt]{4.818pt}{0.400pt}}
\put(1059.0,739.0){\rule[-0.200pt]{4.818pt}{0.400pt}}
\put(1130.0,756.0){\rule[-0.200pt]{0.400pt}{1.927pt}}
\put(1120.0,756.0){\rule[-0.200pt]{4.818pt}{0.400pt}}
\put(1120.0,764.0){\rule[-0.200pt]{4.818pt}{0.400pt}}
\put(1190.0,779.0){\rule[-0.200pt]{0.400pt}{2.168pt}}
\put(1180.0,779.0){\rule[-0.200pt]{4.818pt}{0.400pt}}
\put(1180.0,788.0){\rule[-0.200pt]{4.818pt}{0.400pt}}
\put(1251.0,804.0){\rule[-0.200pt]{0.400pt}{2.168pt}}
\put(1241.0,804.0){\rule[-0.200pt]{4.818pt}{0.400pt}}
\put(1241.0,813.0){\rule[-0.200pt]{4.818pt}{0.400pt}}
\put(1312.0,828.0){\rule[-0.200pt]{0.400pt}{2.409pt}}
\put(1302.0,828.0){\rule[-0.200pt]{4.818pt}{0.400pt}}
\put(1302.0,838.0){\rule[-0.200pt]{4.818pt}{0.400pt}}
\put(1372.0,851.0){\rule[-0.200pt]{0.400pt}{2.650pt}}
\put(1362.0,851.0){\rule[-0.200pt]{4.818pt}{0.400pt}}
\put(1362.0,862.0){\rule[-0.200pt]{4.818pt}{0.400pt}}
\put(1433.0,875.0){\rule[-0.200pt]{0.400pt}{2.650pt}}
\put(1423.0,875.0){\rule[-0.200pt]{4.818pt}{0.400pt}}
\put(1423.0,886.0){\rule[-0.200pt]{4.818pt}{0.400pt}}
\put(1493.0,898.0){\rule[-0.200pt]{0.400pt}{2.891pt}}
\put(1483.0,898.0){\rule[-0.200pt]{4.818pt}{0.400pt}}
\put(1483.0,910.0){\rule[-0.200pt]{4.818pt}{0.400pt}}
\put(1554.0,922.0){\rule[-0.200pt]{0.400pt}{3.132pt}}
\put(1544.0,922.0){\rule[-0.200pt]{4.818pt}{0.400pt}}
\put(1544.0,935.0){\rule[-0.200pt]{4.818pt}{0.400pt}}
\put(1615.0,945.0){\rule[-0.200pt]{0.400pt}{3.132pt}}
\put(1605.0,945.0){\rule[-0.200pt]{4.818pt}{0.400pt}}
\put(1605.0,958.0){\rule[-0.200pt]{4.818pt}{0.400pt}}
\put(1675.0,968.0){\rule[-0.200pt]{0.400pt}{3.373pt}}
\put(1665.0,968.0){\rule[-0.200pt]{4.818pt}{0.400pt}}
\put(1665.0,982.0){\rule[-0.200pt]{4.818pt}{0.400pt}}
\put(1736.0,991.0){\rule[-0.200pt]{0.400pt}{3.613pt}}
\put(1726.0,991.0){\rule[-0.200pt]{4.818pt}{0.400pt}}
\put(1726.0,1006.0){\rule[-0.200pt]{4.818pt}{0.400pt}}
\sbox{\plotpoint}{\rule[-0.400pt]{0.800pt}{0.800pt}}%
\put(220,113){\usebox{\plotpoint}}
\multiput(221.41,113.00)(0.508,1.201){23}{\rule{0.122pt}{2.067pt}}
\multiput(218.34,113.00)(15.000,30.711){2}{\rule{0.800pt}{1.033pt}}
\multiput(236.41,148.00)(0.508,2.794){23}{\rule{0.122pt}{4.467pt}}
\multiput(233.34,148.00)(15.000,70.729){2}{\rule{0.800pt}{2.233pt}}
\multiput(251.41,228.00)(0.508,2.687){23}{\rule{0.122pt}{4.307pt}}
\multiput(248.34,228.00)(15.000,68.061){2}{\rule{0.800pt}{2.153pt}}
\multiput(266.41,305.00)(0.507,1.220){25}{\rule{0.122pt}{2.100pt}}
\multiput(263.34,305.00)(16.000,33.641){2}{\rule{0.800pt}{1.050pt}}
\put(281,340.84){\rule{3.614pt}{0.800pt}}
\multiput(281.00,341.34)(7.500,-1.000){2}{\rule{1.807pt}{0.800pt}}
\multiput(296.00,340.09)(0.493,-0.508){23}{\rule{1.000pt}{0.122pt}}
\multiput(296.00,340.34)(12.924,-15.000){2}{\rule{0.500pt}{0.800pt}}
\multiput(311.00,325.08)(0.863,-0.516){11}{\rule{1.533pt}{0.124pt}}
\multiput(311.00,325.34)(11.817,-9.000){2}{\rule{0.767pt}{0.800pt}}
\put(326,317.34){\rule{3.614pt}{0.800pt}}
\multiput(326.00,316.34)(7.500,2.000){2}{\rule{1.807pt}{0.800pt}}
\multiput(341.00,321.38)(2.104,0.560){3}{\rule{2.600pt}{0.135pt}}
\multiput(341.00,318.34)(9.604,5.000){2}{\rule{1.300pt}{0.800pt}}
\put(356,324.84){\rule{3.854pt}{0.800pt}}
\multiput(356.00,323.34)(8.000,3.000){2}{\rule{1.927pt}{0.800pt}}
\put(372,325.34){\rule{3.614pt}{0.800pt}}
\multiput(372.00,326.34)(7.500,-2.000){2}{\rule{1.807pt}{0.800pt}}
\put(387,323.34){\rule{3.614pt}{0.800pt}}
\multiput(387.00,324.34)(7.500,-2.000){2}{\rule{1.807pt}{0.800pt}}
\put(402,321.84){\rule{3.614pt}{0.800pt}}
\multiput(402.00,322.34)(7.500,-1.000){2}{\rule{1.807pt}{0.800pt}}
\put(432,321.84){\rule{3.614pt}{0.800pt}}
\multiput(432.00,321.34)(7.500,1.000){2}{\rule{1.807pt}{0.800pt}}
\put(417.0,323.0){\rule[-0.400pt]{3.613pt}{0.800pt}}
\put(478,321.84){\rule{3.614pt}{0.800pt}}
\multiput(478.00,322.34)(7.500,-1.000){2}{\rule{1.807pt}{0.800pt}}
\put(493,320.84){\rule{3.614pt}{0.800pt}}
\multiput(493.00,321.34)(7.500,-1.000){2}{\rule{1.807pt}{0.800pt}}
\put(447.0,324.0){\rule[-0.400pt]{7.468pt}{0.800pt}}
\put(554,319.84){\rule{3.614pt}{0.800pt}}
\multiput(554.00,320.34)(7.500,-1.000){2}{\rule{1.807pt}{0.800pt}}
\put(508.0,322.0){\rule[-0.400pt]{11.081pt}{0.800pt}}
\put(584,318.84){\rule{3.614pt}{0.800pt}}
\multiput(584.00,319.34)(7.500,-1.000){2}{\rule{1.807pt}{0.800pt}}
\put(599,317.84){\rule{3.614pt}{0.800pt}}
\multiput(599.00,318.34)(7.500,-1.000){2}{\rule{1.807pt}{0.800pt}}
\put(569.0,321.0){\rule[-0.400pt]{3.613pt}{0.800pt}}
\put(644,316.84){\rule{3.854pt}{0.800pt}}
\multiput(644.00,317.34)(8.000,-1.000){2}{\rule{1.927pt}{0.800pt}}
\put(614.0,319.0){\rule[-0.400pt]{7.227pt}{0.800pt}}
\put(675,315.84){\rule{3.614pt}{0.800pt}}
\multiput(675.00,316.34)(7.500,-1.000){2}{\rule{1.807pt}{0.800pt}}
\put(690,314.84){\rule{3.614pt}{0.800pt}}
\multiput(690.00,315.34)(7.500,-1.000){2}{\rule{1.807pt}{0.800pt}}
\put(660.0,318.0){\rule[-0.400pt]{3.613pt}{0.800pt}}
\put(720,313.84){\rule{3.614pt}{0.800pt}}
\multiput(720.00,314.34)(7.500,-1.000){2}{\rule{1.807pt}{0.800pt}}
\put(705.0,316.0){\rule[-0.400pt]{3.613pt}{0.800pt}}
\put(751,312.84){\rule{3.614pt}{0.800pt}}
\multiput(751.00,313.34)(7.500,-1.000){2}{\rule{1.807pt}{0.800pt}}
\put(766,311.84){\rule{3.614pt}{0.800pt}}
\multiput(766.00,312.34)(7.500,-1.000){2}{\rule{1.807pt}{0.800pt}}
\put(735.0,315.0){\rule[-0.400pt]{3.854pt}{0.800pt}}
\put(796,310.84){\rule{3.614pt}{0.800pt}}
\multiput(796.00,311.34)(7.500,-1.000){2}{\rule{1.807pt}{0.800pt}}
\put(811,309.84){\rule{3.614pt}{0.800pt}}
\multiput(811.00,310.34)(7.500,-1.000){2}{\rule{1.807pt}{0.800pt}}
\put(781.0,313.0){\rule[-0.400pt]{3.613pt}{0.800pt}}
\put(842,308.84){\rule{3.614pt}{0.800pt}}
\multiput(842.00,309.34)(7.500,-1.000){2}{\rule{1.807pt}{0.800pt}}
\put(857,307.84){\rule{3.614pt}{0.800pt}}
\multiput(857.00,308.34)(7.500,-1.000){2}{\rule{1.807pt}{0.800pt}}
\put(872,306.84){\rule{3.614pt}{0.800pt}}
\multiput(872.00,307.34)(7.500,-1.000){2}{\rule{1.807pt}{0.800pt}}
\put(887,305.84){\rule{3.614pt}{0.800pt}}
\multiput(887.00,306.34)(7.500,-1.000){2}{\rule{1.807pt}{0.800pt}}
\put(826.0,311.0){\rule[-0.400pt]{3.854pt}{0.800pt}}
\put(917,304.84){\rule{3.854pt}{0.800pt}}
\multiput(917.00,305.34)(8.000,-1.000){2}{\rule{1.927pt}{0.800pt}}
\put(933,303.84){\rule{3.614pt}{0.800pt}}
\multiput(933.00,304.34)(7.500,-1.000){2}{\rule{1.807pt}{0.800pt}}
\put(948,302.84){\rule{3.614pt}{0.800pt}}
\multiput(948.00,303.34)(7.500,-1.000){2}{\rule{1.807pt}{0.800pt}}
\put(963,301.84){\rule{3.614pt}{0.800pt}}
\multiput(963.00,302.34)(7.500,-1.000){2}{\rule{1.807pt}{0.800pt}}
\put(978,300.84){\rule{3.614pt}{0.800pt}}
\multiput(978.00,301.34)(7.500,-1.000){2}{\rule{1.807pt}{0.800pt}}
\put(993,299.84){\rule{3.614pt}{0.800pt}}
\multiput(993.00,300.34)(7.500,-1.000){2}{\rule{1.807pt}{0.800pt}}
\put(1008,298.84){\rule{3.614pt}{0.800pt}}
\multiput(1008.00,299.34)(7.500,-1.000){2}{\rule{1.807pt}{0.800pt}}
\put(902.0,307.0){\rule[-0.400pt]{3.613pt}{0.800pt}}
\put(1039,297.84){\rule{3.614pt}{0.800pt}}
\multiput(1039.00,298.34)(7.500,-1.000){2}{\rule{1.807pt}{0.800pt}}
\put(1054,296.84){\rule{3.614pt}{0.800pt}}
\multiput(1054.00,297.34)(7.500,-1.000){2}{\rule{1.807pt}{0.800pt}}
\put(1069,295.84){\rule{3.614pt}{0.800pt}}
\multiput(1069.00,296.34)(7.500,-1.000){2}{\rule{1.807pt}{0.800pt}}
\put(1084,294.84){\rule{3.614pt}{0.800pt}}
\multiput(1084.00,295.34)(7.500,-1.000){2}{\rule{1.807pt}{0.800pt}}
\put(1023.0,300.0){\rule[-0.400pt]{3.854pt}{0.800pt}}
\put(1130,294.84){\rule{3.614pt}{0.800pt}}
\multiput(1130.00,294.34)(7.500,1.000){2}{\rule{1.807pt}{0.800pt}}
\put(1145,295.84){\rule{3.614pt}{0.800pt}}
\multiput(1145.00,295.34)(7.500,1.000){2}{\rule{1.807pt}{0.800pt}}
\put(1160,296.84){\rule{3.614pt}{0.800pt}}
\multiput(1160.00,296.34)(7.500,1.000){2}{\rule{1.807pt}{0.800pt}}
\put(1175,298.84){\rule{3.614pt}{0.800pt}}
\multiput(1175.00,297.34)(7.500,3.000){2}{\rule{1.807pt}{0.800pt}}
\put(1190,301.84){\rule{3.614pt}{0.800pt}}
\multiput(1190.00,300.34)(7.500,3.000){2}{\rule{1.807pt}{0.800pt}}
\put(1205,304.84){\rule{3.854pt}{0.800pt}}
\multiput(1205.00,303.34)(8.000,3.000){2}{\rule{1.927pt}{0.800pt}}
\put(1221,308.34){\rule{3.200pt}{0.800pt}}
\multiput(1221.00,306.34)(8.358,4.000){2}{\rule{1.600pt}{0.800pt}}
\put(1236,312.34){\rule{3.200pt}{0.800pt}}
\multiput(1236.00,310.34)(8.358,4.000){2}{\rule{1.600pt}{0.800pt}}
\put(1251,315.84){\rule{3.614pt}{0.800pt}}
\multiput(1251.00,314.34)(7.500,3.000){2}{\rule{1.807pt}{0.800pt}}
\put(1266,318.34){\rule{3.614pt}{0.800pt}}
\multiput(1266.00,317.34)(7.500,2.000){2}{\rule{1.807pt}{0.800pt}}
\put(1281,319.84){\rule{3.614pt}{0.800pt}}
\multiput(1281.00,319.34)(7.500,1.000){2}{\rule{1.807pt}{0.800pt}}
\put(1296,319.84){\rule{3.854pt}{0.800pt}}
\multiput(1296.00,320.34)(8.000,-1.000){2}{\rule{1.927pt}{0.800pt}}
\put(1312,318.34){\rule{3.614pt}{0.800pt}}
\multiput(1312.00,319.34)(7.500,-2.000){2}{\rule{1.807pt}{0.800pt}}
\put(1327,315.34){\rule{3.200pt}{0.800pt}}
\multiput(1327.00,317.34)(8.358,-4.000){2}{\rule{1.600pt}{0.800pt}}
\put(1342,311.34){\rule{3.200pt}{0.800pt}}
\multiput(1342.00,313.34)(8.358,-4.000){2}{\rule{1.600pt}{0.800pt}}
\put(1357,307.34){\rule{3.200pt}{0.800pt}}
\multiput(1357.00,309.34)(8.358,-4.000){2}{\rule{1.600pt}{0.800pt}}
\put(1372,303.84){\rule{3.614pt}{0.800pt}}
\multiput(1372.00,305.34)(7.500,-3.000){2}{\rule{1.807pt}{0.800pt}}
\put(1387,301.34){\rule{3.614pt}{0.800pt}}
\multiput(1387.00,302.34)(7.500,-2.000){2}{\rule{1.807pt}{0.800pt}}
\put(1099.0,296.0){\rule[-0.400pt]{7.468pt}{0.800pt}}
\put(1433,300.84){\rule{3.614pt}{0.800pt}}
\multiput(1433.00,300.34)(7.500,1.000){2}{\rule{1.807pt}{0.800pt}}
\put(1402.0,302.0){\rule[-0.400pt]{7.468pt}{0.800pt}}
\put(1463,300.84){\rule{3.614pt}{0.800pt}}
\multiput(1463.00,301.34)(7.500,-1.000){2}{\rule{1.807pt}{0.800pt}}
\put(1478,299.34){\rule{3.614pt}{0.800pt}}
\multiput(1478.00,300.34)(7.500,-2.000){2}{\rule{1.807pt}{0.800pt}}
\put(1493,297.34){\rule{3.854pt}{0.800pt}}
\multiput(1493.00,298.34)(8.000,-2.000){2}{\rule{1.927pt}{0.800pt}}
\put(1509,295.34){\rule{3.614pt}{0.800pt}}
\multiput(1509.00,296.34)(7.500,-2.000){2}{\rule{1.807pt}{0.800pt}}
\put(1524,293.84){\rule{3.614pt}{0.800pt}}
\multiput(1524.00,294.34)(7.500,-1.000){2}{\rule{1.807pt}{0.800pt}}
\put(1448.0,303.0){\rule[-0.400pt]{3.613pt}{0.800pt}}
\put(1554,294.84){\rule{3.614pt}{0.800pt}}
\multiput(1554.00,293.34)(7.500,3.000){2}{\rule{1.807pt}{0.800pt}}
\put(1569,298.34){\rule{3.200pt}{0.800pt}}
\multiput(1569.00,296.34)(8.358,4.000){2}{\rule{1.600pt}{0.800pt}}
\multiput(1584.00,303.39)(1.579,0.536){5}{\rule{2.333pt}{0.129pt}}
\multiput(1584.00,300.34)(11.157,6.000){2}{\rule{1.167pt}{0.800pt}}
\multiput(1600.00,309.39)(1.467,0.536){5}{\rule{2.200pt}{0.129pt}}
\multiput(1600.00,306.34)(10.434,6.000){2}{\rule{1.100pt}{0.800pt}}
\multiput(1615.00,315.40)(1.176,0.526){7}{\rule{1.914pt}{0.127pt}}
\multiput(1615.00,312.34)(11.027,7.000){2}{\rule{0.957pt}{0.800pt}}
\multiput(1630.00,322.39)(1.467,0.536){5}{\rule{2.200pt}{0.129pt}}
\multiput(1630.00,319.34)(10.434,6.000){2}{\rule{1.100pt}{0.800pt}}
\multiput(1645.00,328.38)(2.104,0.560){3}{\rule{2.600pt}{0.135pt}}
\multiput(1645.00,325.34)(9.604,5.000){2}{\rule{1.300pt}{0.800pt}}
\put(1660,331.84){\rule{3.614pt}{0.800pt}}
\multiput(1660.00,330.34)(7.500,3.000){2}{\rule{1.807pt}{0.800pt}}
\put(1675,333.84){\rule{3.854pt}{0.800pt}}
\multiput(1675.00,333.34)(8.000,1.000){2}{\rule{1.927pt}{0.800pt}}
\put(1691,333.34){\rule{3.614pt}{0.800pt}}
\multiput(1691.00,334.34)(7.500,-2.000){2}{\rule{1.807pt}{0.800pt}}
\put(1706,330.34){\rule{3.200pt}{0.800pt}}
\multiput(1706.00,332.34)(8.358,-4.000){2}{\rule{1.600pt}{0.800pt}}
\multiput(1721.00,328.08)(1.176,-0.526){7}{\rule{1.914pt}{0.127pt}}
\multiput(1721.00,328.34)(11.027,-7.000){2}{\rule{0.957pt}{0.800pt}}
\put(1539.0,295.0){\rule[-0.400pt]{3.613pt}{0.800pt}}
\end{picture}
\caption{An enhanced region of $0\leq t\leq 25$ of Figure 1. The perturbative
estimate of $\la B^2(t)\ra_T$ is shown by the solid curve.}
\label{ill1}
\end{figure}
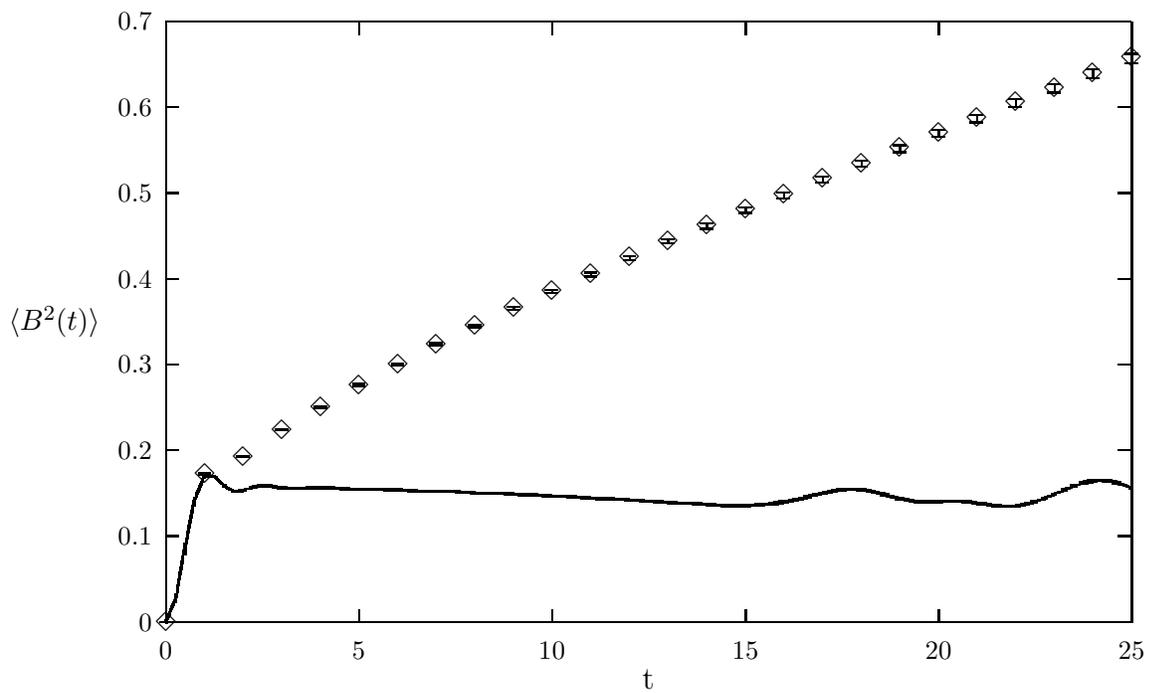

A typical plot of $\langle B^2(t)\rangle_T$ as a function of the lag $t$
is shown in Figures \ref{ill} and \ref{ill1}. As expected, for very short lags
($t\leq 1$)
$\langle B^2(t)\rangle_T$ is mostly accounted for by the perturbative result
obtained from \rf{5.2}. For larger lags a random walk
of $N_{\rm cs}$ sets in, and $\langle B^2(t)\rangle_T$ rapidly approaches a
linear asymptote. Thus, for large enough $t$, $\langle B^2(t)\rangle_T$ can
be fitted to the form (\ref{1.3e}). The range of $t$ suitable for this
linear fit is bounded from below by the initial transient behavior, and from
above by the growing statistical errors in our determination of
$\langle B^2(t)\rangle_T$. With the temperatures and lattice sizes used, we
found the optimal range to be approximately $40<t<200$. As Figures \ref{ill}
and \ref{ill1} indicate, at these lags
$\langle B^2(t)\rangle_T$ is dominated by the nonperturbative effects.
The central values of
$\langle B^2(t)\rangle_T$ for larger $t$ are consistent with linear
dependence on $t$.

\begin{figure}
% GNUPLOT: LaTeX picture
\setlength{\unitlength}{0.240900pt}
\ifx\plotpoint\undefined\newsavebox{\plotpoint}\fi
\sbox{\plotpoint}{\rule[-0.200pt]{0.400pt}{0.400pt}}%
\begin{picture}(1800,1080)(0,0)
\font\gnuplot=cmr10 at 10pt
\gnuplot
\sbox{\plotpoint}{\rule[-0.200pt]{0.400pt}{0.400pt}}%
\put(220.0,113.0){\rule[-0.200pt]{365.204pt}{0.400pt}}
\put(220.0,113.0){\rule[-0.200pt]{0.400pt}{227.410pt}}
\put(220.0,113.0){\rule[-0.200pt]{4.818pt}{0.400pt}}
\put(198,113){\makebox(0,0)[r]{0}}
\put(1716.0,113.0){\rule[-0.200pt]{4.818pt}{0.400pt}}
\put(220.0,231.0){\rule[-0.200pt]{4.818pt}{0.400pt}}
\put(198,231){\makebox(0,0)[r]{0.5}}
\put(1716.0,231.0){\rule[-0.200pt]{4.818pt}{0.400pt}}
\put(220.0,349.0){\rule[-0.200pt]{4.818pt}{0.400pt}}
\put(198,349){\makebox(0,0)[r]{1}}
\put(1716.0,349.0){\rule[-0.200pt]{4.818pt}{0.400pt}}
\put(220.0,467.0){\rule[-0.200pt]{4.818pt}{0.400pt}}
\put(198,467){\makebox(0,0)[r]{1.5}}
\put(1716.0,467.0){\rule[-0.200pt]{4.818pt}{0.400pt}}
\put(220.0,585.0){\rule[-0.200pt]{4.818pt}{0.400pt}}
\put(198,585){\makebox(0,0)[r]{2}}
\put(1716.0,585.0){\rule[-0.200pt]{4.818pt}{0.400pt}}
\put(220.0,703.0){\rule[-0.200pt]{4.818pt}{0.400pt}}
\put(198,703){\makebox(0,0)[r]{2.5}}
\put(1716.0,703.0){\rule[-0.200pt]{4.818pt}{0.400pt}}
\put(220.0,821.0){\rule[-0.200pt]{4.818pt}{0.400pt}}
\put(198,821){\makebox(0,0)[r]{3}}
\put(1716.0,821.0){\rule[-0.200pt]{4.818pt}{0.400pt}}
\put(220.0,939.0){\rule[-0.200pt]{4.818pt}{0.400pt}}
\put(198,939){\makebox(0,0)[r]{3.5}}
\put(1716.0,939.0){\rule[-0.200pt]{4.818pt}{0.400pt}}
\put(220.0,1057.0){\rule[-0.200pt]{4.818pt}{0.400pt}}
\put(198,1057){\makebox(0,0)[r]{4}}
\put(1716.0,1057.0){\rule[-0.200pt]{4.818pt}{0.400pt}}
\put(220.0,113.0){\rule[-0.200pt]{0.400pt}{4.818pt}}
\put(220,68){\makebox(0,0){0}}
\put(220.0,1037.0){\rule[-0.200pt]{0.400pt}{4.818pt}}
\put(599.0,113.0){\rule[-0.200pt]{0.400pt}{4.818pt}}
\put(599,68){\makebox(0,0){50}}
\put(599.0,1037.0){\rule[-0.200pt]{0.400pt}{4.818pt}}
\put(978.0,113.0){\rule[-0.200pt]{0.400pt}{4.818pt}}
\put(978,68){\makebox(0,0){100}}
\put(978.0,1037.0){\rule[-0.200pt]{0.400pt}{4.818pt}}
\put(1357.0,113.0){\rule[-0.200pt]{0.400pt}{4.818pt}}
\put(1357,68){\makebox(0,0){150}}
\put(1357.0,1037.0){\rule[-0.200pt]{0.400pt}{4.818pt}}
\put(1736.0,113.0){\rule[-0.200pt]{0.400pt}{4.818pt}}
\put(1736,68){\makebox(0,0){200}}
\put(1736.0,1037.0){\rule[-0.200pt]{0.400pt}{4.818pt}}
\put(220.0,113.0){\rule[-0.200pt]{365.204pt}{0.400pt}}
\put(1736.0,113.0){\rule[-0.200pt]{0.400pt}{227.410pt}}
\put(220.0,1057.0){\rule[-0.200pt]{365.204pt}{0.400pt}}
\put(45,585){\makebox(0,0){$\langle B^2(t)\rangle$}}
\put(978,23){\makebox(0,0){{\large t}}}
\put(220.0,113.0){\rule[-0.200pt]{0.400pt}{227.410pt}}
\put(220,113){\raisebox{-.8pt}{\makebox(0,0){$\Diamond$}}}
\put(258,178){\raisebox{-.8pt}{\makebox(0,0){$\Diamond$}}}
\put(296,204){\raisebox{-.8pt}{\makebox(0,0){$\Diamond$}}}
\put(334,227){\raisebox{-.8pt}{\makebox(0,0){$\Diamond$}}}
\put(372,248){\raisebox{-.8pt}{\makebox(0,0){$\Diamond$}}}
\put(410,269){\raisebox{-.8pt}{\makebox(0,0){$\Diamond$}}}
\put(447,289){\raisebox{-.8pt}{\makebox(0,0){$\Diamond$}}}
\put(485,310){\raisebox{-.8pt}{\makebox(0,0){$\Diamond$}}}
\put(523,330){\raisebox{-.8pt}{\makebox(0,0){$\Diamond$}}}
\put(561,351){\raisebox{-.8pt}{\makebox(0,0){$\Diamond$}}}
\put(599,372){\raisebox{-.8pt}{\makebox(0,0){$\Diamond$}}}
\put(637,392){\raisebox{-.8pt}{\makebox(0,0){$\Diamond$}}}
\put(675,412){\raisebox{-.8pt}{\makebox(0,0){$\Diamond$}}}
\put(713,431){\raisebox{-.8pt}{\makebox(0,0){$\Diamond$}}}
\put(751,451){\raisebox{-.8pt}{\makebox(0,0){$\Diamond$}}}
\put(789,472){\raisebox{-.8pt}{\makebox(0,0){$\Diamond$}}}
\put(826,492){\raisebox{-.8pt}{\makebox(0,0){$\Diamond$}}}
\put(864,512){\raisebox{-.8pt}{\makebox(0,0){$\Diamond$}}}
\put(902,533){\raisebox{-.8pt}{\makebox(0,0){$\Diamond$}}}
\put(940,553){\raisebox{-.8pt}{\makebox(0,0){$\Diamond$}}}
\put(978,573){\raisebox{-.8pt}{\makebox(0,0){$\Diamond$}}}
\put(1016,594){\raisebox{-.8pt}{\makebox(0,0){$\Diamond$}}}
\put(1054,616){\raisebox{-.8pt}{\makebox(0,0){$\Diamond$}}}
\put(1092,637){\raisebox{-.8pt}{\makebox(0,0){$\Diamond$}}}
\put(1130,659){\raisebox{-.8pt}{\makebox(0,0){$\Diamond$}}}
\put(1168,680){\raisebox{-.8pt}{\makebox(0,0){$\Diamond$}}}
\put(1205,702){\raisebox{-.8pt}{\makebox(0,0){$\Diamond$}}}
\put(1243,723){\raisebox{-.8pt}{\makebox(0,0){$\Diamond$}}}
\put(1281,744){\raisebox{-.8pt}{\makebox(0,0){$\Diamond$}}}
\put(1319,766){\raisebox{-.8pt}{\makebox(0,0){$\Diamond$}}}
\put(1357,788){\raisebox{-.8pt}{\makebox(0,0){$\Diamond$}}}
\put(1395,809){\raisebox{-.8pt}{\makebox(0,0){$\Diamond$}}}
\put(1433,830){\raisebox{-.8pt}{\makebox(0,0){$\Diamond$}}}
\put(1471,851){\raisebox{-.8pt}{\makebox(0,0){$\Diamond$}}}
\put(1509,873){\raisebox{-.8pt}{\makebox(0,0){$\Diamond$}}}
\put(1547,894){\raisebox{-.8pt}{\makebox(0,0){$\Diamond$}}}
\put(1584,915){\raisebox{-.8pt}{\makebox(0,0){$\Diamond$}}}
\put(1622,936){\raisebox{-.8pt}{\makebox(0,0){$\Diamond$}}}
\put(1660,956){\raisebox{-.8pt}{\makebox(0,0){$\Diamond$}}}
\put(1698,978){\raisebox{-.8pt}{\makebox(0,0){$\Diamond$}}}
\put(1736,1000){\raisebox{-.8pt}{\makebox(0,0){$\Diamond$}}}
\put(220,113){\usebox{\plotpoint}}
\put(210.0,113.0){\rule[-0.200pt]{4.818pt}{0.400pt}}
\put(210.0,113.0){\rule[-0.200pt]{4.818pt}{0.400pt}}
\put(258.0,178.0){\usebox{\plotpoint}}
\put(248.0,178.0){\rule[-0.200pt]{4.818pt}{0.400pt}}
\put(248.0,179.0){\rule[-0.200pt]{4.818pt}{0.400pt}}
\put(296.0,204.0){\usebox{\plotpoint}}
\put(286.0,204.0){\rule[-0.200pt]{4.818pt}{0.400pt}}
\put(286.0,205.0){\rule[-0.200pt]{4.818pt}{0.400pt}}
\put(334.0,226.0){\rule[-0.200pt]{0.400pt}{0.482pt}}
\put(324.0,226.0){\rule[-0.200pt]{4.818pt}{0.400pt}}
\put(324.0,228.0){\rule[-0.200pt]{4.818pt}{0.400pt}}
\put(372.0,247.0){\rule[-0.200pt]{0.400pt}{0.482pt}}
\put(362.0,247.0){\rule[-0.200pt]{4.818pt}{0.400pt}}
\put(362.0,249.0){\rule[-0.200pt]{4.818pt}{0.400pt}}
\put(410.0,267.0){\rule[-0.200pt]{0.400pt}{0.723pt}}
\put(400.0,267.0){\rule[-0.200pt]{4.818pt}{0.400pt}}
\put(400.0,270.0){\rule[-0.200pt]{4.818pt}{0.400pt}}
\put(447.0,287.0){\rule[-0.200pt]{0.400pt}{0.964pt}}
\put(437.0,287.0){\rule[-0.200pt]{4.818pt}{0.400pt}}
\put(437.0,291.0){\rule[-0.200pt]{4.818pt}{0.400pt}}
\put(485.0,308.0){\rule[-0.200pt]{0.400pt}{0.964pt}}
\put(475.0,308.0){\rule[-0.200pt]{4.818pt}{0.400pt}}
\put(475.0,312.0){\rule[-0.200pt]{4.818pt}{0.400pt}}
\put(523.0,328.0){\rule[-0.200pt]{0.400pt}{1.204pt}}
\put(513.0,328.0){\rule[-0.200pt]{4.818pt}{0.400pt}}
\put(513.0,333.0){\rule[-0.200pt]{4.818pt}{0.400pt}}
\put(561.0,348.0){\rule[-0.200pt]{0.400pt}{1.686pt}}
\put(551.0,348.0){\rule[-0.200pt]{4.818pt}{0.400pt}}
\put(551.0,355.0){\rule[-0.200pt]{4.818pt}{0.400pt}}
\put(599.0,368.0){\rule[-0.200pt]{0.400pt}{1.686pt}}
\put(589.0,368.0){\rule[-0.200pt]{4.818pt}{0.400pt}}
\put(589.0,375.0){\rule[-0.200pt]{4.818pt}{0.400pt}}
\put(637.0,388.0){\rule[-0.200pt]{0.400pt}{1.927pt}}
\put(627.0,388.0){\rule[-0.200pt]{4.818pt}{0.400pt}}
\put(627.0,396.0){\rule[-0.200pt]{4.818pt}{0.400pt}}
\put(675.0,407.0){\rule[-0.200pt]{0.400pt}{2.409pt}}
\put(665.0,407.0){\rule[-0.200pt]{4.818pt}{0.400pt}}
\put(665.0,417.0){\rule[-0.200pt]{4.818pt}{0.400pt}}
\put(713.0,426.0){\rule[-0.200pt]{0.400pt}{2.650pt}}
\put(703.0,426.0){\rule[-0.200pt]{4.818pt}{0.400pt}}
\put(703.0,437.0){\rule[-0.200pt]{4.818pt}{0.400pt}}
\put(751.0,445.0){\rule[-0.200pt]{0.400pt}{2.891pt}}
\put(741.0,445.0){\rule[-0.200pt]{4.818pt}{0.400pt}}
\put(741.0,457.0){\rule[-0.200pt]{4.818pt}{0.400pt}}
\put(789.0,465.0){\rule[-0.200pt]{0.400pt}{3.132pt}}
\put(779.0,465.0){\rule[-0.200pt]{4.818pt}{0.400pt}}
\put(779.0,478.0){\rule[-0.200pt]{4.818pt}{0.400pt}}
\put(826.0,485.0){\rule[-0.200pt]{0.400pt}{3.373pt}}
\put(816.0,485.0){\rule[-0.200pt]{4.818pt}{0.400pt}}
\put(816.0,499.0){\rule[-0.200pt]{4.818pt}{0.400pt}}
\put(864.0,504.0){\rule[-0.200pt]{0.400pt}{3.854pt}}
\put(854.0,504.0){\rule[-0.200pt]{4.818pt}{0.400pt}}
\put(854.0,520.0){\rule[-0.200pt]{4.818pt}{0.400pt}}
\put(902.0,524.0){\rule[-0.200pt]{0.400pt}{4.095pt}}
\put(892.0,524.0){\rule[-0.200pt]{4.818pt}{0.400pt}}
\put(892.0,541.0){\rule[-0.200pt]{4.818pt}{0.400pt}}
\put(940.0,544.0){\rule[-0.200pt]{0.400pt}{4.336pt}}
\put(930.0,544.0){\rule[-0.200pt]{4.818pt}{0.400pt}}
\put(930.0,562.0){\rule[-0.200pt]{4.818pt}{0.400pt}}
\put(978.0,564.0){\rule[-0.200pt]{0.400pt}{4.577pt}}
\put(968.0,564.0){\rule[-0.200pt]{4.818pt}{0.400pt}}
\put(968.0,583.0){\rule[-0.200pt]{4.818pt}{0.400pt}}
\put(1016.0,584.0){\rule[-0.200pt]{0.400pt}{5.059pt}}
\put(1006.0,584.0){\rule[-0.200pt]{4.818pt}{0.400pt}}
\put(1006.0,605.0){\rule[-0.200pt]{4.818pt}{0.400pt}}
\put(1054.0,604.0){\rule[-0.200pt]{0.400pt}{5.541pt}}
\put(1044.0,604.0){\rule[-0.200pt]{4.818pt}{0.400pt}}
\put(1044.0,627.0){\rule[-0.200pt]{4.818pt}{0.400pt}}
\put(1092.0,625.0){\rule[-0.200pt]{0.400pt}{6.022pt}}
\put(1082.0,625.0){\rule[-0.200pt]{4.818pt}{0.400pt}}
\put(1082.0,650.0){\rule[-0.200pt]{4.818pt}{0.400pt}}
\put(1130.0,646.0){\rule[-0.200pt]{0.400pt}{6.263pt}}
\put(1120.0,646.0){\rule[-0.200pt]{4.818pt}{0.400pt}}
\put(1120.0,672.0){\rule[-0.200pt]{4.818pt}{0.400pt}}
\put(1168.0,666.0){\rule[-0.200pt]{0.400pt}{6.986pt}}
\put(1158.0,666.0){\rule[-0.200pt]{4.818pt}{0.400pt}}
\put(1158.0,695.0){\rule[-0.200pt]{4.818pt}{0.400pt}}
\put(1205.0,686.0){\rule[-0.200pt]{0.400pt}{7.468pt}}
\put(1195.0,686.0){\rule[-0.200pt]{4.818pt}{0.400pt}}
\put(1195.0,717.0){\rule[-0.200pt]{4.818pt}{0.400pt}}
\put(1243.0,707.0){\rule[-0.200pt]{0.400pt}{7.709pt}}
\put(1233.0,707.0){\rule[-0.200pt]{4.818pt}{0.400pt}}
\put(1233.0,739.0){\rule[-0.200pt]{4.818pt}{0.400pt}}
\put(1281.0,727.0){\rule[-0.200pt]{0.400pt}{8.431pt}}
\put(1271.0,727.0){\rule[-0.200pt]{4.818pt}{0.400pt}}
\put(1271.0,762.0){\rule[-0.200pt]{4.818pt}{0.400pt}}
\put(1319.0,748.0){\rule[-0.200pt]{0.400pt}{8.913pt}}
\put(1309.0,748.0){\rule[-0.200pt]{4.818pt}{0.400pt}}
\put(1309.0,785.0){\rule[-0.200pt]{4.818pt}{0.400pt}}
\put(1357.0,768.0){\rule[-0.200pt]{0.400pt}{9.395pt}}
\put(1347.0,768.0){\rule[-0.200pt]{4.818pt}{0.400pt}}
\put(1347.0,807.0){\rule[-0.200pt]{4.818pt}{0.400pt}}
\put(1395.0,788.0){\rule[-0.200pt]{0.400pt}{10.118pt}}
\put(1385.0,788.0){\rule[-0.200pt]{4.818pt}{0.400pt}}
\put(1385.0,830.0){\rule[-0.200pt]{4.818pt}{0.400pt}}
\put(1433.0,808.0){\rule[-0.200pt]{0.400pt}{10.600pt}}
\put(1423.0,808.0){\rule[-0.200pt]{4.818pt}{0.400pt}}
\put(1423.0,852.0){\rule[-0.200pt]{4.818pt}{0.400pt}}
\put(1471.0,828.0){\rule[-0.200pt]{0.400pt}{11.081pt}}
\put(1461.0,828.0){\rule[-0.200pt]{4.818pt}{0.400pt}}
\put(1461.0,874.0){\rule[-0.200pt]{4.818pt}{0.400pt}}
\put(1509.0,848.0){\rule[-0.200pt]{0.400pt}{11.804pt}}
\put(1499.0,848.0){\rule[-0.200pt]{4.818pt}{0.400pt}}
\put(1499.0,897.0){\rule[-0.200pt]{4.818pt}{0.400pt}}
\put(1547.0,869.0){\rule[-0.200pt]{0.400pt}{12.045pt}}
\put(1537.0,869.0){\rule[-0.200pt]{4.818pt}{0.400pt}}
\put(1537.0,919.0){\rule[-0.200pt]{4.818pt}{0.400pt}}
\put(1584.0,888.0){\rule[-0.200pt]{0.400pt}{12.768pt}}
\put(1574.0,888.0){\rule[-0.200pt]{4.818pt}{0.400pt}}
\put(1574.0,941.0){\rule[-0.200pt]{4.818pt}{0.400pt}}
\put(1622.0,908.0){\rule[-0.200pt]{0.400pt}{13.249pt}}
\put(1612.0,908.0){\rule[-0.200pt]{4.818pt}{0.400pt}}
\put(1612.0,963.0){\rule[-0.200pt]{4.818pt}{0.400pt}}
\put(1660.0,928.0){\rule[-0.200pt]{0.400pt}{13.731pt}}
\put(1650.0,928.0){\rule[-0.200pt]{4.818pt}{0.400pt}}
\put(1650.0,985.0){\rule[-0.200pt]{4.818pt}{0.400pt}}
\put(1698.0,948.0){\rule[-0.200pt]{0.400pt}{14.454pt}}
\put(1688.0,948.0){\rule[-0.200pt]{4.818pt}{0.400pt}}
\put(1688.0,1008.0){\rule[-0.200pt]{4.818pt}{0.400pt}}
\put(1736.0,968.0){\rule[-0.200pt]{0.400pt}{15.177pt}}
\put(1726.0,968.0){\rule[-0.200pt]{4.818pt}{0.400pt}}
\put(1726.0,1031.0){\rule[-0.200pt]{4.818pt}{0.400pt}}
\end{picture}
\caption{Diffusion of $N_{\rm cs}$: mean square of $B(t)$ as a function of
the lag $t$ for $\beta=12$, $L_L=32$.}
\label{ill}
\end{figure}
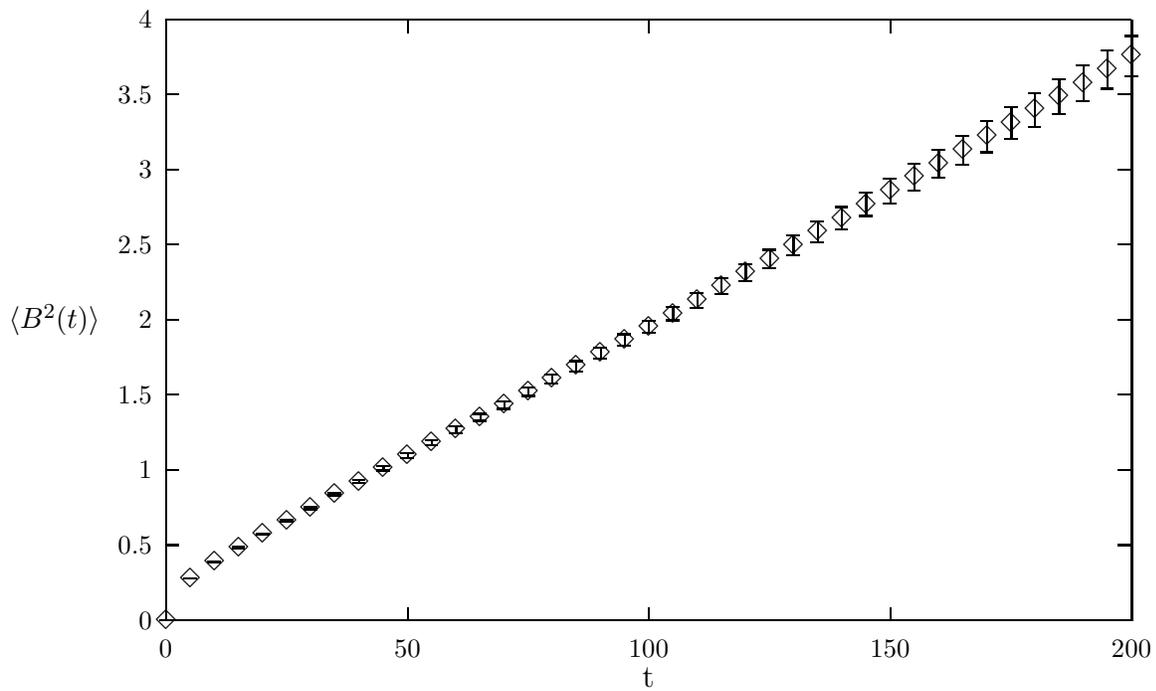

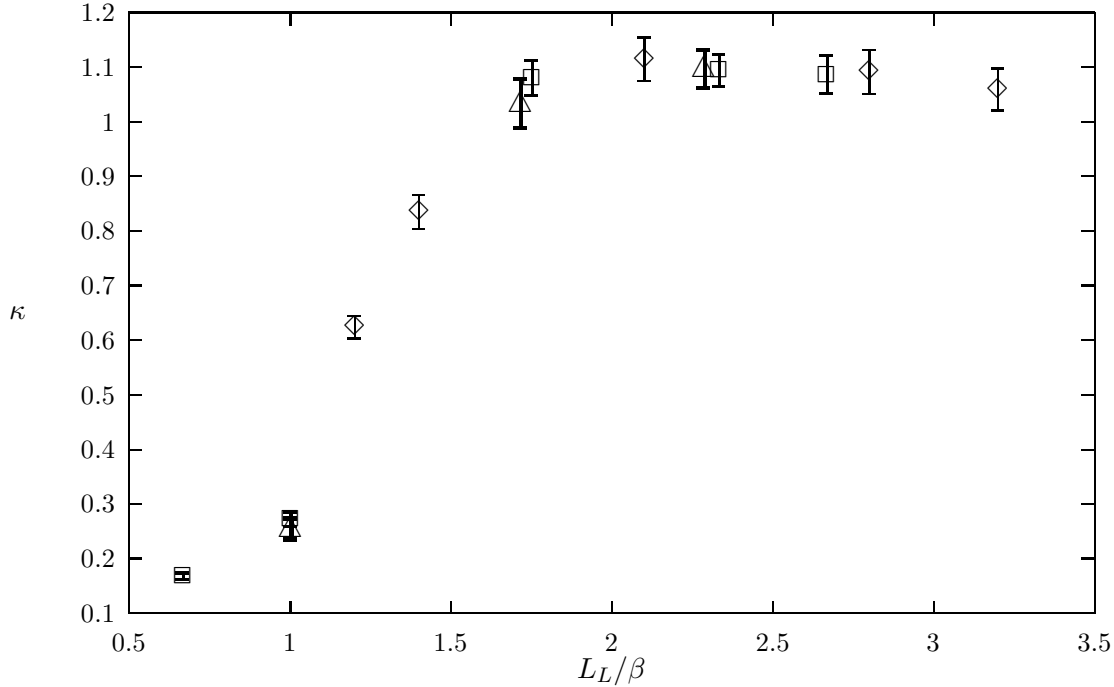
\begin{figure}
% GNUPLOT: LaTeX picture
\setlength{\unitlength}{0.240900pt}
\ifx\plotpoint\undefined\newsavebox{\plotpoint}\fi
\sbox{\plotpoint}{\rule[-0.200pt]{0.400pt}{0.400pt}}%
\begin{picture}(1800,1080)(0,0)
\font\gnuplot=cmr10 at 10pt
\gnuplot
\sbox{\plotpoint}{\rule[-0.200pt]{0.400pt}{0.400pt}}%
\put(220.0,113.0){\rule[-0.200pt]{4.818pt}{0.400pt}}
\put(198,113){\makebox(0,0)[r]{0.1}}
\put(1716.0,113.0){\rule[-0.200pt]{4.818pt}{0.400pt}}
\put(220.0,199.0){\rule[-0.200pt]{4.818pt}{0.400pt}}
\put(198,199){\makebox(0,0)[r]{0.2}}
\put(1716.0,199.0){\rule[-0.200pt]{4.818pt}{0.400pt}}
\put(220.0,285.0){\rule[-0.200pt]{4.818pt}{0.400pt}}
\put(198,285){\makebox(0,0)[r]{0.3}}
\put(1716.0,285.0){\rule[-0.200pt]{4.818pt}{0.400pt}}
\put(220.0,370.0){\rule[-0.200pt]{4.818pt}{0.400pt}}
\put(198,370){\makebox(0,0)[r]{0.4}}
\put(1716.0,370.0){\rule[-0.200pt]{4.818pt}{0.400pt}}
\put(220.0,456.0){\rule[-0.200pt]{4.818pt}{0.400pt}}
\put(198,456){\makebox(0,0)[r]{0.5}}
\put(1716.0,456.0){\rule[-0.200pt]{4.818pt}{0.400pt}}
\put(220.0,542.0){\rule[-0.200pt]{4.818pt}{0.400pt}}
\put(198,542){\makebox(0,0)[r]{0.6}}
\put(1716.0,542.0){\rule[-0.200pt]{4.818pt}{0.400pt}}
\put(220.0,628.0){\rule[-0.200pt]{4.818pt}{0.400pt}}
\put(198,628){\makebox(0,0)[r]{0.7}}
\put(1716.0,628.0){\rule[-0.200pt]{4.818pt}{0.400pt}}
\put(220.0,714.0){\rule[-0.200pt]{4.818pt}{0.400pt}}
\put(198,714){\makebox(0,0)[r]{0.8}}
\put(1716.0,714.0){\rule[-0.200pt]{4.818pt}{0.400pt}}
\put(220.0,800.0){\rule[-0.200pt]{4.818pt}{0.400pt}}
\put(198,800){\makebox(0,0)[r]{0.9}}
\put(1716.0,800.0){\rule[-0.200pt]{4.818pt}{0.400pt}}
\put(220.0,885.0){\rule[-0.200pt]{4.818pt}{0.400pt}}
\put(198,885){\makebox(0,0)[r]{1}}
\put(1716.0,885.0){\rule[-0.200pt]{4.818pt}{0.400pt}}
\put(220.0,971.0){\rule[-0.200pt]{4.818pt}{0.400pt}}
\put(198,971){\makebox(0,0)[r]{1.1}}
\put(1716.0,971.0){\rule[-0.200pt]{4.818pt}{0.400pt}}
\put(220.0,1057.0){\rule[-0.200pt]{4.818pt}{0.400pt}}
\put(198,1057){\makebox(0,0)[r]{1.2}}
\put(1716.0,1057.0){\rule[-0.200pt]{4.818pt}{0.400pt}}
\put(220.0,113.0){\rule[-0.200pt]{0.400pt}{4.818pt}}
\put(220,68){\makebox(0,0){0.5}}
\put(220.0,1037.0){\rule[-0.200pt]{0.400pt}{4.818pt}}
\put(473.0,113.0){\rule[-0.200pt]{0.400pt}{4.818pt}}
\put(473,68){\makebox(0,0){1}}
\put(473.0,1037.0){\rule[-0.200pt]{0.400pt}{4.818pt}}
\put(725.0,113.0){\rule[-0.200pt]{0.400pt}{4.818pt}}
\put(725,68){\makebox(0,0){1.5}}
\put(725.0,1037.0){\rule[-0.200pt]{0.400pt}{4.818pt}}
\put(978.0,113.0){\rule[-0.200pt]{0.400pt}{4.818pt}}
\put(978,68){\makebox(0,0){2}}
\put(978.0,1037.0){\rule[-0.200pt]{0.400pt}{4.818pt}}
\put(1231.0,113.0){\rule[-0.200pt]{0.400pt}{4.818pt}}
\put(1231,68){\makebox(0,0){2.5}}
\put(1231.0,1037.0){\rule[-0.200pt]{0.400pt}{4.818pt}}
\put(1483.0,113.0){\rule[-0.200pt]{0.400pt}{4.818pt}}
\put(1483,68){\makebox(0,0){3}}
\put(1483.0,1037.0){\rule[-0.200pt]{0.400pt}{4.818pt}}
\put(1736.0,113.0){\rule[-0.200pt]{0.400pt}{4.818pt}}
\put(1736,68){\makebox(0,0){3.5}}
\put(1736.0,1037.0){\rule[-0.200pt]{0.400pt}{4.818pt}}
\put(220.0,113.0){\rule[-0.200pt]{365.204pt}{0.400pt}}
\put(1736.0,113.0){\rule[-0.200pt]{0.400pt}{227.410pt}}
\put(220.0,1057.0){\rule[-0.200pt]{365.204pt}{0.400pt}}
\put(45,585){\makebox(0,0){$\kappa$}}
\put(978,23){\makebox(0,0){$L_L/\beta$}}
\put(220.0,113.0){\rule[-0.200pt]{0.400pt}{227.410pt}}
\put(1584,936){\raisebox{-.8pt}{\makebox(0,0){$\Diamond$}}}
\put(1382,964){\raisebox{-.8pt}{\makebox(0,0){$\Diamond$}}}
\put(1029,984){\raisebox{-.8pt}{\makebox(0,0){$\Diamond$}}}
\put(675,744){\raisebox{-.8pt}{\makebox(0,0){$\Diamond$}}}
\put(574,563){\raisebox{-.8pt}{\makebox(0,0){$\Diamond$}}}
\put(1584.0,903.0){\rule[-0.200pt]{0.400pt}{15.899pt}}
\put(1574.0,903.0){\rule[-0.200pt]{4.818pt}{0.400pt}}
\put(1574.0,969.0){\rule[-0.200pt]{4.818pt}{0.400pt}}
\put(1382.0,929.0){\rule[-0.200pt]{0.400pt}{16.622pt}}
\put(1372.0,929.0){\rule[-0.200pt]{4.818pt}{0.400pt}}
\put(1372.0,998.0){\rule[-0.200pt]{4.818pt}{0.400pt}}
\put(1029.0,949.0){\rule[-0.200pt]{0.400pt}{16.622pt}}
\put(1019.0,949.0){\rule[-0.200pt]{4.818pt}{0.400pt}}
\put(1019.0,1018.0){\rule[-0.200pt]{4.818pt}{0.400pt}}
\put(675.0,717.0){\rule[-0.200pt]{0.400pt}{12.768pt}}
\put(665.0,717.0){\rule[-0.200pt]{4.818pt}{0.400pt}}
\put(665.0,770.0){\rule[-0.200pt]{4.818pt}{0.400pt}}
\put(574.0,545.0){\rule[-0.200pt]{0.400pt}{8.431pt}}
\put(564.0,545.0){\rule[-0.200pt]{4.818pt}{0.400pt}}
\put(564.0,580.0){\rule[-0.200pt]{4.818pt}{0.400pt}}
\sbox{\plotpoint}{\rule[-0.400pt]{0.800pt}{0.800pt}}%
\put(1315,959){\raisebox{-.8pt}{\makebox(0,0){$\Box$}}}
\put(1146,966){\raisebox{-.8pt}{\makebox(0,0){$\Box$}}}
\put(852,954){\raisebox{-.8pt}{\makebox(0,0){$\Box$}}}
\put(473,260){\raisebox{-.8pt}{\makebox(0,0){$\Box$}}}
\put(304,171){\raisebox{-.8pt}{\makebox(0,0){$\Box$}}}
\put(1315.0,929.0){\rule[-0.400pt]{0.800pt}{14.454pt}}
\put(1305.0,929.0){\rule[-0.400pt]{4.818pt}{0.800pt}}
\put(1305.0,989.0){\rule[-0.400pt]{4.818pt}{0.800pt}}
\put(1146.0,940.0){\rule[-0.400pt]{0.800pt}{12.286pt}}
\put(1136.0,940.0){\rule[-0.400pt]{4.818pt}{0.800pt}}
\put(1136.0,991.0){\rule[-0.400pt]{4.818pt}{0.800pt}}
\put(852.0,926.0){\rule[-0.400pt]{0.800pt}{13.490pt}}
\put(842.0,926.0){\rule[-0.400pt]{4.818pt}{0.800pt}}
\put(842.0,982.0){\rule[-0.400pt]{4.818pt}{0.800pt}}
\put(473.0,249.0){\rule[-0.400pt]{0.800pt}{5.300pt}}
\put(463.0,249.0){\rule[-0.400pt]{4.818pt}{0.800pt}}
\put(463.0,271.0){\rule[-0.400pt]{4.818pt}{0.800pt}}
\put(304.0,166.0){\rule[-0.400pt]{0.800pt}{2.409pt}}
\put(294.0,166.0){\rule[-0.400pt]{4.818pt}{0.800pt}}
\put(294.0,176.0){\rule[-0.400pt]{4.818pt}{0.800pt}}
\sbox{\plotpoint}{\rule[-0.500pt]{1.000pt}{1.000pt}}%
\sbox{\plotpoint}{\rule[-0.600pt]{1.200pt}{1.200pt}}%
\put(1122,968){\makebox(0,0){$\triangle$}}
\put(834,913){\makebox(0,0){$\triangle$}}
\put(473,245){\makebox(0,0){$\triangle$}}
\put(1122.0,939.0){\rule[-0.600pt]{1.200pt}{14.213pt}}
\put(1112.0,939.0){\rule[-0.600pt]{4.818pt}{1.200pt}}
\put(1112.0,998.0){\rule[-0.600pt]{4.818pt}{1.200pt}}
\put(834.0,875.0){\rule[-0.600pt]{1.200pt}{18.549pt}}
\put(824.0,875.0){\rule[-0.600pt]{4.818pt}{1.200pt}}
\put(824.0,952.0){\rule[-0.600pt]{4.818pt}{1.200pt}}
\put(473.0,229.0){\rule[-0.600pt]{1.200pt}{7.950pt}}
\put(463.0,229.0){\rule[-0.600pt]{4.818pt}{1.200pt}}
\put(463.0,262.0){\rule[-0.600pt]{4.818pt}{1.200pt}}
\end{picture}
\caption{The rate prefactor $\kappa$ dependence on the dimensionless parameter
$L_L/\beta$. The data points are for $\beta=10$ (diamonds), 12 (squares), and
14 (triangles).}
\label{fs}
\end{figure}

We determined the coefficient $\k$ of \rf{1.1y} by performing a correlated
fit of $\langle B^2(t)\rangle_T$ to the form (\ref{1.3e}) for data points
within the optimal range. As Figure~\ref{fs} shows, the values
of $\k$ for different
combinations of $\beta$ and $L_L$ depend, to a good approximation, on a
dimensionless ratio $L_L/\beta$ (note, in particular, the equality of
$\k$ values for $\beta=12,L_L=12$ and for $\beta=14,L_L=14$).
We conclude that in the range of temperatures
considered $\k$ is nearly insensitive to the lattice spacing, {\it i.e.,}
close to a finite continuum limit. On the other hand, $\k$ exhibits strong
finite-size effects and grows rapidly for $L_L/\beta\leq 2$. This is in
agreement with our expectation that the diffusion of $N_{\rm cs}$ is dominated
by large-scale structures whose linear size is of order $\beta$. Once the
lattice is too small to accommodate objects of this size, the rate rapidly
decreases\footnote{That the sphaleron-like transitions are related
to large scale objects has been verified directly by ``freezing'' the
time evolution in the neighborhood of $N_{cs} =n+1/2$. One finds energy
profiles corresponding to extended object \cite{af},
in contrast to the situation when one is far from $N_{cs} =n+1/2$.
In addition one could check that level crossing occurred as expected
from continuum physics.}. For large $L_L/\beta>2$ $\k$ reaches a constant
value.
In fact, the length scale at which $\k$ saturates can be naturally interpreted
in terms of magnetic mass, $m_m=0.466 g^2 T$ in physical units \cite{karsch}:
$\k$ stops growing as soon as the lattice size exceeds twice the inverse
magnetic mass (the factor of 2 relating the two lengths is to be expected
with periodic boundary conditions). Our estimate,
based on $\beta=12,L_L=32$ is $\k=1.09\pm 0.04$. Estimates from other
large-volume measurements are consistent with this one within the error bars.

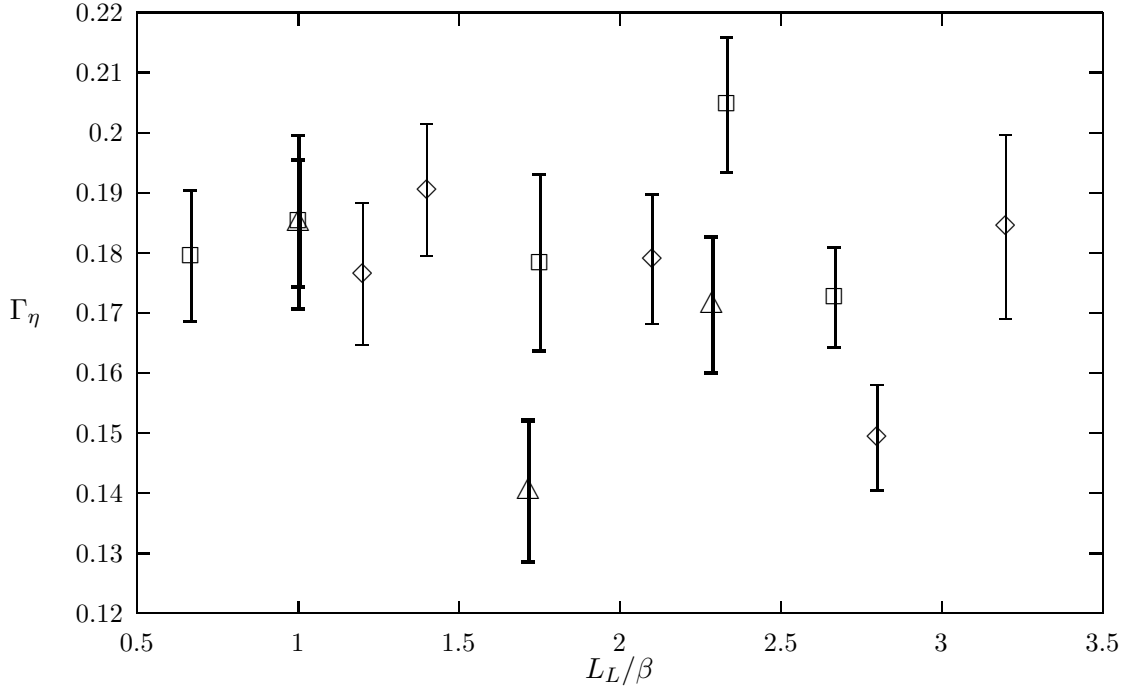
\begin{figure}
% GNUPLOT: LaTeX picture
\setlength{\unitlength}{0.240900pt}
\ifx\plotpoint\undefined\newsavebox{\plotpoint}\fi
\sbox{\plotpoint}{\rule[-0.200pt]{0.400pt}{0.400pt}}%
\begin{picture}(1800,1080)(0,0)
\font\gnuplot=cmr10 at 10pt
\gnuplot
\sbox{\plotpoint}{\rule[-0.200pt]{0.400pt}{0.400pt}}%
\put(220.0,113.0){\rule[-0.200pt]{4.818pt}{0.400pt}}
\put(198,113){\makebox(0,0)[r]{0.12}}
\put(1716.0,113.0){\rule[-0.200pt]{4.818pt}{0.400pt}}
\put(220.0,207.0){\rule[-0.200pt]{4.818pt}{0.400pt}}
\put(198,207){\makebox(0,0)[r]{0.13}}
\put(1716.0,207.0){\rule[-0.200pt]{4.818pt}{0.400pt}}
\put(220.0,302.0){\rule[-0.200pt]{4.818pt}{0.400pt}}
\put(198,302){\makebox(0,0)[r]{0.14}}
\put(1716.0,302.0){\rule[-0.200pt]{4.818pt}{0.400pt}}
\put(220.0,396.0){\rule[-0.200pt]{4.818pt}{0.400pt}}
\put(198,396){\makebox(0,0)[r]{0.15}}
\put(1716.0,396.0){\rule[-0.200pt]{4.818pt}{0.400pt}}
\put(220.0,491.0){\rule[-0.200pt]{4.818pt}{0.400pt}}
\put(198,491){\makebox(0,0)[r]{0.16}}
\put(1716.0,491.0){\rule[-0.200pt]{4.818pt}{0.400pt}}
\put(220.0,585.0){\rule[-0.200pt]{4.818pt}{0.400pt}}
\put(198,585){\makebox(0,0)[r]{0.17}}
\put(1716.0,585.0){\rule[-0.200pt]{4.818pt}{0.400pt}}
\put(220.0,679.0){\rule[-0.200pt]{4.818pt}{0.400pt}}
\put(198,679){\makebox(0,0)[r]{0.18}}
\put(1716.0,679.0){\rule[-0.200pt]{4.818pt}{0.400pt}}
\put(220.0,774.0){\rule[-0.200pt]{4.818pt}{0.400pt}}
\put(198,774){\makebox(0,0)[r]{0.19}}
\put(1716.0,774.0){\rule[-0.200pt]{4.818pt}{0.400pt}}
\put(220.0,868.0){\rule[-0.200pt]{4.818pt}{0.400pt}}
\put(198,868){\makebox(0,0)[r]{0.2}}
\put(1716.0,868.0){\rule[-0.200pt]{4.818pt}{0.400pt}}
\put(220.0,963.0){\rule[-0.200pt]{4.818pt}{0.400pt}}
\put(198,963){\makebox(0,0)[r]{0.21}}
\put(1716.0,963.0){\rule[-0.200pt]{4.818pt}{0.400pt}}
\put(220.0,1057.0){\rule[-0.200pt]{4.818pt}{0.400pt}}
\put(198,1057){\makebox(0,0)[r]{0.22}}
\put(1716.0,1057.0){\rule[-0.200pt]{4.818pt}{0.400pt}}
\put(220.0,113.0){\rule[-0.200pt]{0.400pt}{4.818pt}}
\put(220,68){\makebox(0,0){0.5}}
\put(220.0,1037.0){\rule[-0.200pt]{0.400pt}{4.818pt}}
\put(473.0,113.0){\rule[-0.200pt]{0.400pt}{4.818pt}}
\put(473,68){\makebox(0,0){1}}
\put(473.0,1037.0){\rule[-0.200pt]{0.400pt}{4.818pt}}
\put(725.0,113.0){\rule[-0.200pt]{0.400pt}{4.818pt}}
\put(725,68){\makebox(0,0){1.5}}
\put(725.0,1037.0){\rule[-0.200pt]{0.400pt}{4.818pt}}
\put(978.0,113.0){\rule[-0.200pt]{0.400pt}{4.818pt}}
\put(978,68){\makebox(0,0){2}}
\put(978.0,1037.0){\rule[-0.200pt]{0.400pt}{4.818pt}}
\put(1231.0,113.0){\rule[-0.200pt]{0.400pt}{4.818pt}}
\put(1231,68){\makebox(0,0){2.5}}
\put(1231.0,1037.0){\rule[-0.200pt]{0.400pt}{4.818pt}}
\put(1483.0,113.0){\rule[-0.200pt]{0.400pt}{4.818pt}}
\put(1483,68){\makebox(0,0){3}}
\put(1483.0,1037.0){\rule[-0.200pt]{0.400pt}{4.818pt}}
\put(1736.0,113.0){\rule[-0.200pt]{0.400pt}{4.818pt}}
\put(1736,68){\makebox(0,0){3.5}}
\put(1736.0,1037.0){\rule[-0.200pt]{0.400pt}{4.818pt}}
\put(220.0,113.0){\rule[-0.200pt]{365.204pt}{0.400pt}}
\put(1736.0,113.0){\rule[-0.200pt]{0.400pt}{227.410pt}}
\put(220.0,1057.0){\rule[-0.200pt]{365.204pt}{0.400pt}}
\put(45,585){\makebox(0,0){$\Gamma_\eta$}}
\put(978,23){\makebox(0,0){$L_L/\beta$}}
\put(220.0,113.0){\rule[-0.200pt]{0.400pt}{227.410pt}}
\put(574,646){\raisebox{-.8pt}{\makebox(0,0){$\Diamond$}}}
\put(675,778){\raisebox{-.8pt}{\makebox(0,0){$\Diamond$}}}
\put(1029,669){\raisebox{-.8pt}{\makebox(0,0){$\Diamond$}}}
\put(1382,389){\raisebox{-.8pt}{\makebox(0,0){$\Diamond$}}}
\put(1584,721){\raisebox{-.8pt}{\makebox(0,0){$\Diamond$}}}
\put(574.0,535.0){\rule[-0.200pt]{0.400pt}{53.721pt}}
\put(564.0,535.0){\rule[-0.200pt]{4.818pt}{0.400pt}}
\put(564.0,758.0){\rule[-0.200pt]{4.818pt}{0.400pt}}
\put(675.0,675.0){\rule[-0.200pt]{0.400pt}{49.866pt}}
\put(665.0,675.0){\rule[-0.200pt]{4.818pt}{0.400pt}}
\put(665.0,882.0){\rule[-0.200pt]{4.818pt}{0.400pt}}
\put(1029.0,568.0){\rule[-0.200pt]{0.400pt}{48.903pt}}
\put(1019.0,568.0){\rule[-0.200pt]{4.818pt}{0.400pt}}
\put(1019.0,771.0){\rule[-0.200pt]{4.818pt}{0.400pt}}
\put(1382.0,306.0){\rule[-0.200pt]{0.400pt}{39.989pt}}
\put(1372.0,306.0){\rule[-0.200pt]{4.818pt}{0.400pt}}
\put(1372.0,472.0){\rule[-0.200pt]{4.818pt}{0.400pt}}
\put(1584.0,576.0){\rule[-0.200pt]{0.400pt}{69.620pt}}
\put(1574.0,576.0){\rule[-0.200pt]{4.818pt}{0.400pt}}
\put(1574.0,865.0){\rule[-0.200pt]{4.818pt}{0.400pt}}
\sbox{\plotpoint}{\rule[-0.400pt]{0.800pt}{0.800pt}}%
\put(304,674){\raisebox{-.8pt}{\makebox(0,0){$\Box$}}}
\put(473,728){\raisebox{-.8pt}{\makebox(0,0){$\Box$}}}
\put(852,663){\raisebox{-.8pt}{\makebox(0,0){$\Box$}}}
\put(1146,912){\raisebox{-.8pt}{\makebox(0,0){$\Box$}}}
\put(1315,609){\raisebox{-.8pt}{\makebox(0,0){$\Box$}}}
\put(304.0,571.0){\rule[-0.400pt]{0.800pt}{49.625pt}}
\put(294.0,571.0){\rule[-0.400pt]{4.818pt}{0.800pt}}
\put(294.0,777.0){\rule[-0.400pt]{4.818pt}{0.800pt}}
\put(473.0,591.0){\rule[-0.400pt]{0.800pt}{65.766pt}}
\put(463.0,591.0){\rule[-0.400pt]{4.818pt}{0.800pt}}
\put(463.0,864.0){\rule[-0.400pt]{4.818pt}{0.800pt}}
\put(852.0,525.0){\rule[-0.400pt]{0.800pt}{66.729pt}}
\put(842.0,525.0){\rule[-0.400pt]{4.818pt}{0.800pt}}
\put(842.0,802.0){\rule[-0.400pt]{4.818pt}{0.800pt}}
\put(1146.0,806.0){\rule[-0.400pt]{0.800pt}{50.830pt}}
\put(1136.0,806.0){\rule[-0.400pt]{4.818pt}{0.800pt}}
\put(1136.0,1017.0){\rule[-0.400pt]{4.818pt}{0.800pt}}
\put(1315.0,531.0){\rule[-0.400pt]{0.800pt}{37.821pt}}
\put(1305.0,531.0){\rule[-0.400pt]{4.818pt}{0.800pt}}
\put(1305.0,688.0){\rule[-0.400pt]{4.818pt}{0.800pt}}
\sbox{\plotpoint}{\rule[-0.500pt]{1.000pt}{1.000pt}}%
\sbox{\plotpoint}{\rule[-0.600pt]{1.200pt}{1.200pt}}%
\put(473,725){\makebox(0,0){$\triangle$}}
\put(834,304){\makebox(0,0){$\triangle$}}
\put(1122,597){\makebox(0,0){$\triangle$}}
\put(473.0,626.0){\rule[-0.600pt]{1.200pt}{47.939pt}}
\put(463.0,626.0){\rule[-0.600pt]{4.818pt}{1.200pt}}
\put(463.0,825.0){\rule[-0.600pt]{4.818pt}{1.200pt}}
\put(834.0,193.0){\rule[-0.600pt]{1.200pt}{53.721pt}}
\put(824.0,193.0){\rule[-0.600pt]{4.818pt}{1.200pt}}
\put(824.0,416.0){\rule[-0.600pt]{4.818pt}{1.200pt}}
\put(1122.0,490.0){\rule[-0.600pt]{1.200pt}{51.553pt}}
\put(1112.0,490.0){\rule[-0.600pt]{4.818pt}{1.200pt}}
\put(1112.0,704.0){\rule[-0.600pt]{4.818pt}{1.200pt}}
\end{picture}
\caption{Diffusion constant per unit volume $\Gamma_\eta$ of $\eta(t)$ plotted
against the dimensionless parameter $L_L/\beta$.
The data points are for $\beta=10$ (diamonds), 12 (squares), and
14 (triangles).}
\label{visc}
\end{figure}

Our results for $\k$ are in sharp contrast with those for the shear viscosity,
extracted from the diffusion rate of $\eta(t)$. In the chosen range of
temperatures $\eta(t)$ behaves as an underdamped random walk, and determination
of the corresponding diffusion constant requires knowledge of
$\langle\eta^2(t)\rangle_T$ at lags $t$ of the order of 1000 (as compared to
$t\leq 200$ for the Chern-Simons number), making it a difficult task.
As Figure~\ref{visc} shows, the corresponding diffusion rate $\Gamma_\eta$
shows no marked temperature or finite-size dependence. The absence of finite
size effects for $\Gamma_\eta$ indicates that, unlike $\k$, this quantity is
not predominantly sensitive to long-wavelength modes of the system.
Independence
of $\Gamma_\eta$ of the temperature means,
on dimensional grounds, that $\Gamma_\eta$, and hence the classical shear
viscosity,
diverges as $a^{-4}$, where $a$ is the lattice spacing. This cutoff
dependence of transport coefficients is found in a similar situation in
condensed-matter physics, for phonons at temperatures far above
the Debye temperature \cite{berman}.

\section{Discussion}

We have seen that $\k \approx 1$ in a classical $SU(2)$ gauge theory,
and that we consequently have an observable (the diffusion rate
between gauge equivalent non-Abelian vacua) which is
not plagued by the generic ultraviolet thermal
fluctuations of classical field theory. In addition we expect that
the result obtained describes the high temperature limit of the
classical field theory corresponding to the standard model.

While this result is interesting by itself, it is of even more
interest to understand to what extent it is a good approximation
to the full quantum theory. In the Introduction we gave
qualitative arguments in favor of reasonable agreement between the
classical result for the rate and the full quantum one, essentially reproducing
the considerations which originally led Grigoriev, Rubakov
and Shaposhnikov to suggest the classical real time method.
Unfortunately, not much is known at present about the size and functional form
of quantum corrections to dynamical, real-time quantities.
The best we can offer is an educated guess based on a much better understood
relation between the static properties of the quantum theory and of the
effective dimensinally reduced one \cite{bielefeld,fks}.
The full thermal path integral,
where the bosonic fields obey periodic boundary conditions in the
imaginary time direction,
reduces in the high temperature limit to a partition function of a
three-dimensional statistical mechanical problem.
This partition function inherits the $UV$ thermal cutoff
from the underlying quantum field theory. If a lattice regularization
is used in the quantum theory, the spatial part of this lattice will provide
the regularization of the thermal three-dimensional theory.
It is known that dimensional reduction is not completely equivalent to the
classical approximation.
The correspondence between the two is conveniently described in terms
of the field component $A_0 (x)$ which is the Lagrange multiplier
of the Gauss' law. The latter may be imposed with the help of a
$\del$-function factor in the classical statistical weight. Alternatively,
one can introduce $A_0 (x)$ and integrate out the color electric fields. The
classical partition function then takes form
\beq{4.1}
Z = \int \cD A_i \cD A_0 \; e^{-S_{{\rm eff}} (A_i,A_0)},
\eeq
where
\beq{4.2}
S_{{\rm eff}} (A)  =\frac{1}{4g^2 T}\int d^3 x\; \left[ F^a_{ij}F^a_{ij}
+ (D_i^{ab} A_0^b)^2\right].
\eeq
On the other hand, the dimensionally reduced partition function has the same
form as \rf{4.1}, but this time the effective action includes Higgs potential
terms for $A_0$ and reads
\beq{4.3}
S_{{\rm eff}} (A) = \frac{1}{4g^2(T) T}\int d^3 x \;\left[F^a_{ij}F^a_{ij}
+ (D_i^{ab} A_0^b)^2 +\m(T) (A^a_0)^2 + c g(T)^2 ((A_0^a)^2)^2 +\cdots
\right],
\eeq
where the parameters $g(T)$ and $\m(T)$ are calculable functions
of the temperature, $g(T)$ being the running, temperature
dependent coupling constant, unlike in the purely classical case.
The important point is that, despite the apparent inequivalence of the two
partition functions,
we do not expect any qualitative difference
between the  dimensionally reduced theory given by \rf{4.3} and the classical
one given by \rf{4.2}, since they both are in the confined
phase. In particular, both theories should give a magnetic screening
mass of order $g^2 T$. It therefore seems likely that the classical theory
is able to generate the approximatively correct non-perturbative
behavior at the magnetic-mass scale, and that the replacement $\a_w \to\a_w
(T)$
in the classical expression \rf{1.1y} yields a reasonable transition rate
estimate in the full theory.

\section*{Acknowledgments}
We thank L. McLerran and M.E. Shaposhnikov for useful comments.
This work was supported by the Danish Research Council under contract
No. 9500713. The bulk of the simulations was perfomed on the Cray-C92 and
the SP2 supercomputers at UNI-C.

\section*{Appendix}
In this Appendix we briefly describe thermal properties of
Chern-Simons number in the noncompact classical U(1) theory whose lattice
Hamiltonian is
\beq{5.1}
H\equiv {1\over 2}\sum_{j,n}E^2_{j,n}
+{1\over 4}\sum_{j,m\neq n}\left(\theta_{j,n}+\theta_{j+n,m}-
\theta_{j+m,n}-\theta_{j,m}\right)^2,\eeq
where, in the usual lattice notation, $j$ are sites and $m,n$ are positive
directions on a cubic lattice, and $\theta$ are lattice gauge potentials,
whose conjugate momenta $E_{j,n}$ are the electric fields.
Note that upon substitution $U_{j,n}=\exp(i\sigma^\alpha\theta^\alpha_{j,n})$
and linearization \rf{2.1} reduces to \rf{5.1} summed over
color indices $\alpha$. Imposing the Coulomb gauge
$$\sum_n\left(\theta_{j,n}-\theta_{j-n,n}\right)=0,$$
we eliminate longitudinal degrees of freedom (the electric field is transversal
by the Gauss' law). In momentum representation \rf{5.1} becomes
$$H={1\over 2}\sum_p \left[{\cal E}^i_p {\cal E}^i_{-p}
+4\sum_m\sin^2\left({p_m\over 2}\right)a^i_p a^i_{-p}\right],$$
where the Cartesian components of the momentum are $p_m=2\pi l_m/L_L$ with
integer
$1-L_L/2\leq l_m\leq L_L/2$ on an $L_L^3$ lattice with periodic boundary
conditions,
$a^i_{-p}=(a^i_p)^*$ is a photon field with momentum $p$ and polarization $i$,
and ${\cal E}^i_p=({\cal E}^i_{-p})^*$ are electric fields in momentum space.
It is easy to see that a mode with momentum $p$ has an eigenfrequency
$$\omega_p=2\sqrt{\sum_m\sin^2\left({p_m\over 2}\right)}.$$
In the same notation the Chern-Simons number is
\begin{eqnarray}
N_{\rm cs}&=&{1\over{32\pi^2}}\epsilon_{lmn}\sum_j\left(\theta_{j-n,n}+
\theta_{j,n}
\right)\left(\theta_{j-l+m,l}+\theta_{j+m,l}-\theta_{j-m,l}-\theta_{j-l-m,l}
\right)\nonumber\\
&=&{1\over{4\pi^2i}}\epsilon_{lmn}\sum_p {\rm e}^{i(p_l-p_n)/2}
\cos\left({p_l\over 2}\right)\cos\left({p_n\over 2}\right)
\sin(p_m)e^i(p)_n e^j(-p)_l a_p^i a_{-p}^j,\label{nlat}\end{eqnarray}
where the transverse polarization vectors $e^i(p)$ on the lattice have the
properties
$$e^i(-p)=\left(e^{i}(p)\right)^*; \ \ e^i(p)_n e^j(-p)_n=\delta^{ij};
\ \ e^i(p)_l e^i(-p)_n=\delta_{ln}-{\rm e}^{{i\over 2}(p_l-p_n)}
{{\sin\left({p_l\over 2}\right)\sin\left({p_n\over 2}\right)}
\over{\sum_r\sin^2\left({p_r\over 2}\right)}}.$$
Using these properties, and performing a straightforward Gaussian integration
with the thermal weight $\exp(-\beta H)$, one obtains the autocorrelation
function
\beq{5.2}
\la N_{\rm cs}(t)N_{\rm cs}(0)\ra_T=
{1\over{4\pi^4\beta^2}}\sum_p{{\cos^2\left(\omega_pt\right)}\over{\omega^2_p}}
\prod_l\cos^2\left({p_l\over 2}\right). \eeq
This expression can be explicitly evaluated for any finite lattice. The thermal
width $\nu$ of Abelian $N_{\rm cs}$ is given by \rf{5.2} at $t=0$. In the
limit $L_L\to\infty$ the sum in \rf{5.2} may be replaced by an integral, and
$\nu$ becomes an extensive quantity: $\nu\approx 6.930\times
10^{-4}\beta^{-2}L_L^3$. We used this estimate, together with the conversion
rule \rf{2.7}, to obtain \rf{1.3dd}. Note that the long-time behavior of
\rf{5.2} depends on the order in which the thermodynamic $L_L\to\infty$
and the $t\to\infty$ limits are taken: if the latter is taken first (which is
the case of practical interest here) on a finite-size lattice, \rf{5.2} will
at long times fluctuate around $\nu$ and not approach zero,
as can be verified by taking time average of \rf{5.2} over a long time
interval.

Another useful Abelian quantity is the thermal width of $\partial_t N_{\rm
cs}$.
Taking time derivative of (\ref{nlat}) and averaging over the thermal ensemble
yields
\beq{5.3}
\la \dot N^2_{\rm cs}\ra_T={1\over{2\pi^4\beta^2}}\sum_p
\prod_l\cos^2\left({p_l\over 2}\right) \to {L_L^3\over{16\pi^4\beta^2}},\eeq
where in the last step the large-volume limit was taken.

\end{document}